\documentclass[sigconf]{acmart}

\AtBeginDocument{%
  \providecommand\BibTeX{{%
    \normalfont B\kern-0.5em{\scshape i\kern-0.25em b}\kern-0.8em\TeX}}}

\usepackage{algorithm}
\usepackage{algorithmic}
\usepackage{caption}
\usepackage{subcaption}
\usepackage{makecell}

\usepackage{multirow}
\setcopyright{acmcopyright}
\copyrightyear{2018}
\acmYear{2018}
\acmDOI{XXXXXXX.XXXXXXX}

\acmConference[Conference acronym 'XX]{Make sure to enter the correct
  conference title from your rights confirmation emai}{June 03--05,
  2018}{Woodstock, NY}
\acmPrice{15.00}
\acmISBN{978-1-4503-XXXX-X/18/06}

\begin{document}

\title{Select and Trade: Towards Unified Pair Trading with Hierarchical Reinforcement Learning}

\author{Weiguang Han}
\email{han.wei.guang@whu.edu.cn
}
\author{Boyi Zhang}
\email{zhangby@whu.edu.cn
}
\author{Min Peng}
\email{pengm@whu.edu.cn
}
\affiliation{%
  \institution{Wuhan University}
  \city{Wuhan}
  \state{Hubei}
  \country{China}
  \postcode{430072}
}
\author{Qianqian Xie}
\email{qianqian.xie@manchester.ac.uk
}
\affiliation{%
  \institution{University of Manchester}
  \city{Manchester}
  \country{United Kingdom}
}

\author{Yanzhao Lai}
\email{laiyanzhao@swjtu.edu.cn
}
\affiliation{%
  \institution{Southwest Jiaotong University}
  \city{Chengdu}
  \state{Sichuan}
  \country{China}
}
\author{Jimin Huang}
\authornotemark[1]
\email{jimin@chancefocus.com
}
\affiliation{%
  \institution{Chancefocus AMC.}
  \city{Shanghai}
  \country{China}
}

\renewcommand{\shortauthors}{Han, et al.}

\begin{abstract}
    Pair trading is one of the most effective statistical arbitrage strategies which seeks a neutral profit by hedging a pair of selected assets. Existing methods generally decompose the task into two separate steps: pair selection and trading. However, the decoupling of two closely related sub-tasks can block information propagation and lead to limited overall performance. For pair selection, ignoring the trading performance results in the wrong assets being selected with irrelevant price movements, while the agent trained for trading can overfit to the selected assets without any historical information of other assets. To address it, in this paper, we propose a paradigm for automatic pair trading as a unified task rather than a two-step pipeline. We design a hierarchical reinforcement learning framework to jointly learn and optimize two sub-tasks. A high-level policy would select two assets from all possible combinations and a low-level policy would then perform a series of trading actions. Experimental results on real-world stock data demonstrate the effectiveness of our method on pair trading compared with both existing pair selection and trading methods.
\end{abstract}

	


\received{20 February 2007}
\received[revised]{12 March 2009}
\received[accepted]{5 June 2009}

\maketitle

\section{Introduction}
Since 1987, pair trading, a basic statistical arbitrage approach, has been intensively practised and studied~\cite{Lehoczky2018OverviewAH}.
It is an integral component of the financial market and plays a crucial role in enhancing market efficiency~\cite{Shearer2021StabilityEO}.
On worldwide markets and varied asset types, such as stocks, futures, and cryptocurrency, it has been argued that pair trading is effective in the long term~\cite{Krauss2017}.
In contrast to portfolio selection to find the optimal portfolio with the highest ``risky profit''~\cite{Hunanyan2019PortfolioS}, it seeks ``riskless profit'' by performing arbitrage tradings on the abnormal price movements of two correlated assets~\cite{Lehoczky2018OverviewAH}.
It first picks two correlated assets and monitors the spread between their respective prices. If the spread widens abnormally, it will execute trading operations on two assets and gain a profit when the spread recovers to its usual value. For instance, if Google's price is usually \$2 higher than Facebook's and it suddenly rises to \$5 higher, the strategy will short Google (expect its price to fall) and long Facebook (expect its price to increase) and close two tradings when the spread returns to \$2. The total return of the strategy is the sum of the returns from the two transactions on Google and Facebook, which relies solely on the spread between the two assets. Therefore, it is irrelevant to the vast majority of common risks, such as market fluctuations, since the profit resulting from the risk on Google would compensate for the loss resulting from the risk on Facebook. Nonetheless, it depends on two crucial factors: (1) the chosen two assets should be beneficial for pair trading, with a spread that exhibits significant mean-reversion and high volatility; and (2) a flexible agent that can identify abnormal increases and falls of spread from normal fluctuations.
	
Generally, previous methods for pair trading divided the process into two discrete stages: \textbf{pair selection} and \textbf{trading}. For pair selection, they generally employ predefined statistical tests or fundamental distance measurements to select two assets based on their historical price ~\cite{gatev2006pairs,vidyamurthy2004pairs,Huck2015PairsTA,Do2012ArePT,Chen2019EmpiricalIO,Pole2007StatisticalAA,Perlin2007MOA,Rad2015ThePO,Lin2006LossPI,Puspaningrum2010FindingTO,elliott2005pairs,Galenko2012TradingIT,Bertram2009AnalyticSF,Chen2014PairsTV}.
For example, a number of previous researches apply the cointegration test~\cite{vidyamurthy2004pairs} to determine if the historical price spread between two assets is stable.
After selecting the pair, they would engage in trading using fixed-threshold-based strategies to generate the return in a subsequent period.
Recently, inspired by the successful deployment of reinforcement learning in other areas~\cite{fischer2018reinforcement,almahdi2019constrained,katongo2021use,lucarelli2019deep}, there have been efforts to introduce reinforcement learning to train a flexible agent and report a significant improvement over traditional methods~\cite{fallahpour2016pairs, Kim2022HybridDR, wang2021improving}.
	
However, existing methods of automated pair trading are still confronted with drawbacks.
Despite the fact that they can ensure the relevance between the selected two assets and perform tradings by decoupling pair selection and trading,
it prevents the flow of information between them, which can be a significant concern since they are tightly coupled.
For pair selection, existing methods would choose the wrong asset pairs since the employed model-free metrics are \textbf{target irrelevant}, which means, they consider no performance of candidate asset pairs during the following trading period.
For example, the optimal asset pair with the lowest Euclidean distance would have zero spread and trading opportunities.
It is fundamental to dynamically learn the measurement of the future profitability of asset pairs from the data.
As for trading, existing methods can be \textbf{target overfitting} due to only observing the pre-selected asset pair during the training and ignoring other asset pairs and the market.
Although reinforcement learning allows their methods to learn a flexible agent which can explore different trading actions during the training, the learned agent could show poor performance in the trading period with unseen market data since only partial historical information is leveraged.

Despite the critical necessity to jointly simulate the two phases of pair trading, there have been no prior attempts in this area. In this research, we propose a novel paradigm for automated pair trading, in which the two-step process is formulated as a unified task rather than a pipeline with two independently sorted sub-tasks. The approach must simultaneously choose the trading pair from candidate pairs in a formation period and trade it in a later trading period in order to optimize trading performance.
Although the paradigm is straightforward for the task, it poses two challenges to the development of successful approaches.
First, it is challenging to represent the sequential process of the paradigm in which trading occurs after pair selection, i.e., selecting two correlated assets and then trading on their anomalies. 
Second, there are complicated relationships between pair selection and trading that must be fully utilized to generate risk-free profits. 
Pair selection intuitively sets the input of trading, while trading offers the output of pair selection in the form of profit.

To address these issues, we design a new framework \textbf{TRIALS} that adopts feudal hierarchical reinforcement learning (FHRL) \cite{Pateria2021HierarchicalRL} to jointly learn and optimize two steps: a high-level reinforcement learning policy as the manager for pair selection, and a low-level reinforcement learning policy as the worker for trading.
The agent in our proposed framework would first select an asset pair from all possible combinations of assets, and then perform a series of trading actions based on the selected pair.
Given a set of assets,
for the high-level manager, \textbf{states} are the historical price features of these assets in the formation period; \textbf{options} are all possible combinations of these assets; \textbf{rewards} are the overall performance which is generated from the low-level worker on the trading period.
As For the low-level worker, given the chosen option as two selected assets, i.e, Google and Facebook, \textbf{states} are the historical price features of two assets and the trading information of the agent such as historical actions, cash, and present net value; \textbf{actions} are three discrete trading actions including \textit{long} (Buy Google to sell it later and sell Facebook to buy it back later), \textit{short} (Sell Google and buy Facebook), and \textit{clear} (Sale previous bought assets and buy previously sold assets to close tradings); and \textbf{rewards} are the overall performance of the agent in the formation period.
Notice that the rewards for the high-level manager and the low-level worker are devised on the trading and formation period respectively, although they are both generated via the same low-level worker.
This allows our method to optimize the agent at two levels jointly, which guides the high-level manager to select optimal asset pairs according to their trading performance on the unseen market data, and forces the low-level worker to consider different asset pairs and capture the common profitable patterns for pair trading.
We further verify the effectiveness of our method on U.S. and Chinese stock datasets compared with both previous pair selection and trading methods.
The experimental results prove that our proposed method can achieve the best performance, which attributes to the correct pair selection and corresponding precise trading actions.
	
In summary, our contributions can be listed as:
\begin{enumerate}
\item We are the first to introduce a new task for pair trading that combines the existing two tasks as pair selection and trading. In order to optimize the total trading performance, it is necessary for the approach to simultaneously consider these two steps, which were previously overlooked in both pair selection and trading.
\item We design a novel end-to-end hierarchical framework that introduces feudal hierarchical reinforcement learning to jointly optimize a high-level policy for pair selection and a low-level policy for trading.
\item Experimental results on both U.S. and Chinese stock markets demonstrate the effectiveness of our method compared with existing pair selection and trading methods.
\end{enumerate}

\section{Related Work}
\subsection{Traditional Pair Selection}
For pair selection, previous methods aim to find two assets whose prices have moved together historically in a formation period, and their future spread is assumed to be historical mean-reverted~\cite{Krauss2017}.
They generally adopted statistical or fundamental similarity measurements based on historical price information to perform asset pair selection before trading.
The distance approach was first introduced~\cite{gatev2006pairs,Huck2015PairsTA,Do2012ArePT,Chen2019EmpiricalIO,Pole2007StatisticalAA,Perlin2007MOA} for pair selection, which simply adopted distance metrics such as the sum of Euclidean squared distance (SSD) for the price time series to model the connection between two assets.
However, an ideal asset pair in these model-free methods were expected to be two assets with exactly the same price movement in historical time, which have zero trading opportunities for no fluctuations of price spread.
There were also methods~\cite{vidyamurthy2004pairs,Rad2015ThePO,Lin2006LossPI,Puspaningrum2010FindingTO,Elliott2005PairsT,Galenko2012TradingIT,Bertram2009AnalyticSF,Chen2014PairsTV} that directly model the tradability of a candidate pair based on the Engle-Granger cointegration test, which performs linear regression using the price series of two assets and expects the residual to be stationary.

However, the mean-reversion properties of the spread of an asset pair in the future can be irrelevant to their mean-reversion strength in history, which limits the trading performance of the selected pair from these parameter-free methods.
Although there were also methods \cite{Krauss2017DeepNN} that integrated neural networks to learn the metrics, they proposed to measure the profit of assets rather than the asset pairs and selected the top and bottom assets to form the trading pair, which is difficult to find two matched assets.
\cite{Xu2020DynamicPM} was the most similar study which considers pair trading as a unified portfolio management task.
Their methods using historical price spread as the metric nevertheless suffer from the same issue, even though they can dynamically learn the trading and allocation ratios of each pair.

\subsection{Reinforcement Learning for Pair Trading}
After pair selection, previous methods generate trading signals which trigger contradictory actions on two assets during the trading period.
Based on the assumption that the spread of the selected pair would still revert to its historical mean value, previous methods generally employ simple threshold-based rules that they would long the undervalued and short the overvalued asset when the spread is higher or lower than the historical mean by pre-defined thresholds~\cite{Krauss2017}.
However, it requires expert knowledge to identify the optimal trading thresholds in the time-varying market.

Inspired by the success of applying reinforcement learning (RL) in financial trading problems \cite{fischer2018reinforcement}, 
previous attempts generally focused on introducing RL methods to develop flexible trading agents after pair selection via traditional methods.
\cite{fallahpour2016pairs} used the cointegration method to select trading pairs, and adopted Q-Learning \cite{watkins1992q} to select optimal trading parameters.
\citeauthor{kim2019optimizing} introduced a deep Q-network \cite{mnih2015human} to select the best trading threshold for cointegration approaches \cite{kim2019optimizing}.
\cite{Lu2022StructuralBP} proposed to detect structural changes and improve reinforcement learning trading methods.
\citeauthor{wang2021improving,Brim2020DeepRL} directly utilized the RL methods to train an agent for trading \cite{wang2021improving,Brim2020DeepRL}.
\cite{Kim2022HybridDR} further introduced stop-loss boundaries to control the risk.
Although these methods have shown the benefits of the integration of RLs as a smart trading agent, they still adopt traditional methods for pair selection which only consider the historical performance of the trading pair.
Moreover, their trading agent can easily overfit to the only observable asset pair and show limited performance on the unseen future market.
However, there were no previous efforts to address the problem, which requires the method to jointly learn how to select and trade asset pairs.

\subsection{Hierarchical Reinforcement Learning}
Many approaches have been proposed for building agents within the context of hierarchical reinforcement learning (HRL)~\cite{Xie2021HierarchicalRL, Pope2021HierarchicalRL, Saleh2020HierarchicalRL}. The feudal framework is one popular approach for HRL, in which the action space of a higher-level policy consists of sub-goals corresponding to various sub-tasks and the objective of this lower-level policy is to achieve the input sub-goal \cite{NIPS1992_d14220ee}. In HRL, different levels of temporal abstraction enable efficient credit assignment over longer timescales \cite{Vezhnevets2017FeUdalNF}. At the same time, a subtask may itself be easier to learn and the learned sub-tasks lead to more structured exploration over the course of training of the HRL agent \cite{Nachum2019WhyDH}. In previous works, the low-level policy generally learned handcrafted sub-goals \cite{Kulkarni2016HierarchicalDR}, discovered options \cite{Bacon2017TheOA} or intrinsic rewards \cite{Vezhnevets2017FeUdalNF}, while the high-level policy is learned using extrinsic rewards from the environment.
The decomposition of feudal HRL can also help to model complex tasks that are difficult for normal RL methods.

As one of the most challenging applications, pair trading consisting of two separate steps requires the method to optimize two related but different sub-tasks.
Existing methods generally deem the process as a two-step pipeline and apply different methods for each step respectively.
It inevitably blocks the information propagation between these two steps and introduces extra noise due to the error accumulation step-by-step.
To the best of our knowledge, our work is the first one that applies HRL in pair trading to end-to-end learning and inference.
	
\section{Hierarchical Pair Trading Framework}
\begin{figure*}[htb]
 \centering
 \includegraphics[width=\textwidth]{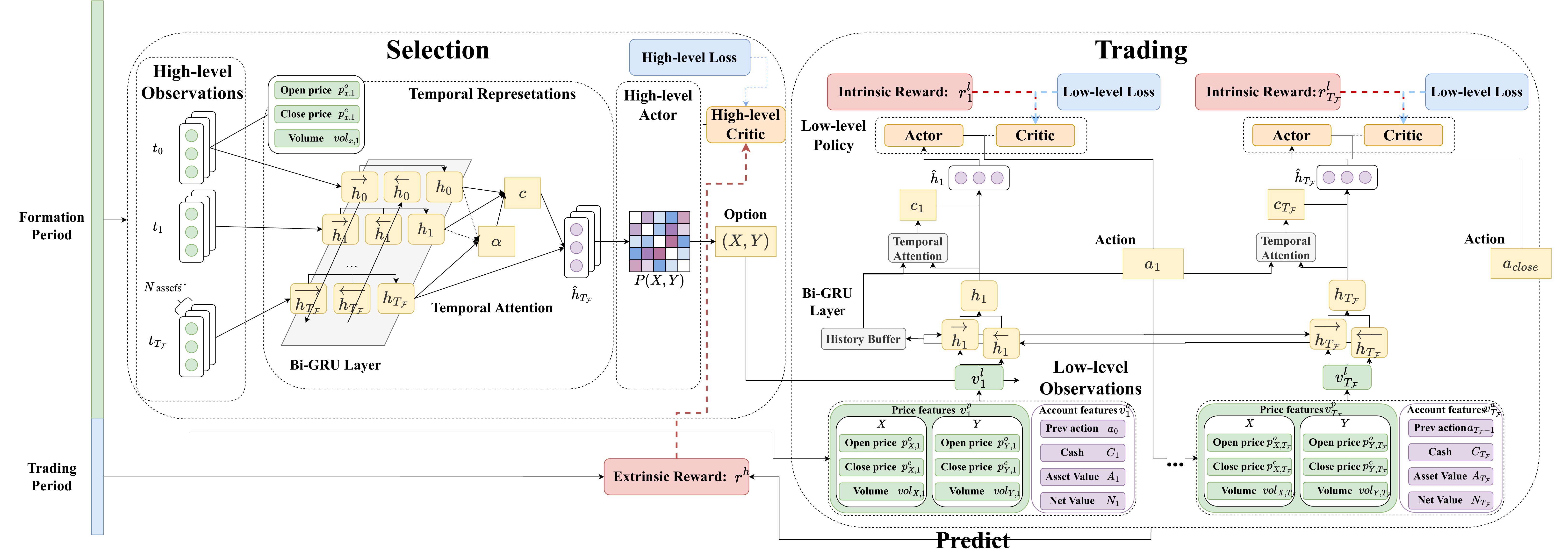}
 \caption{The hierarchical framework for pair trading.}
 \label{fig:framework}
\end{figure*}

In this section, we illustrate the detail of our proposed hierarchical pair trading framework, as shown in Fig. \ref{fig:framework}.
	
\subsection{Formalization}
Generally, pair trading consists of two steps: pair selection and trading.
In pair selection, it would select two correlated assets from all possible combinations of assets to form a trading pair.
Given the trading pair, it would perform a series of trading actions to earn market-neutral profit in a subsequent period.
The task aims to maximize the trading profit of the selected asset pair, which requires selecting the optimal trading pair and choosing correct trading actions during the trading period.
Different from previous approaches that generally take two steps separately, in this paper, we propose to jointly learn to select and trade the pair in a unified hierarchical framework.
Therefore, given a formation period with $T_\mathcal{F}$ time points consisting of $\{0,1,\dots,T_\mathcal{F}-1\}$, a subsequent trading period with $T_\mathcal{T}$ time points consisting of $\{0,1,\dots,T_\mathcal{T}-1\}$, and selected $N$ assets $\mathcal{X} = \{x_1, x_2,\dots,x_{N}\}$,
there are formation price series $\{p_0^x, p_1^x,\dots,p_{T_\mathcal{F}-1}^x\}$ and trading price series $\{p_0^x, p_1^x,\dots,p_{T_\mathcal{T}-1}^x\}$ for each asset $x \in \mathcal{X}$ that is associated with each time point in formation period and trading period respectively.

\begin{figure}[htb]
 \centering
 \includegraphics[width=0.5\columnwidth]{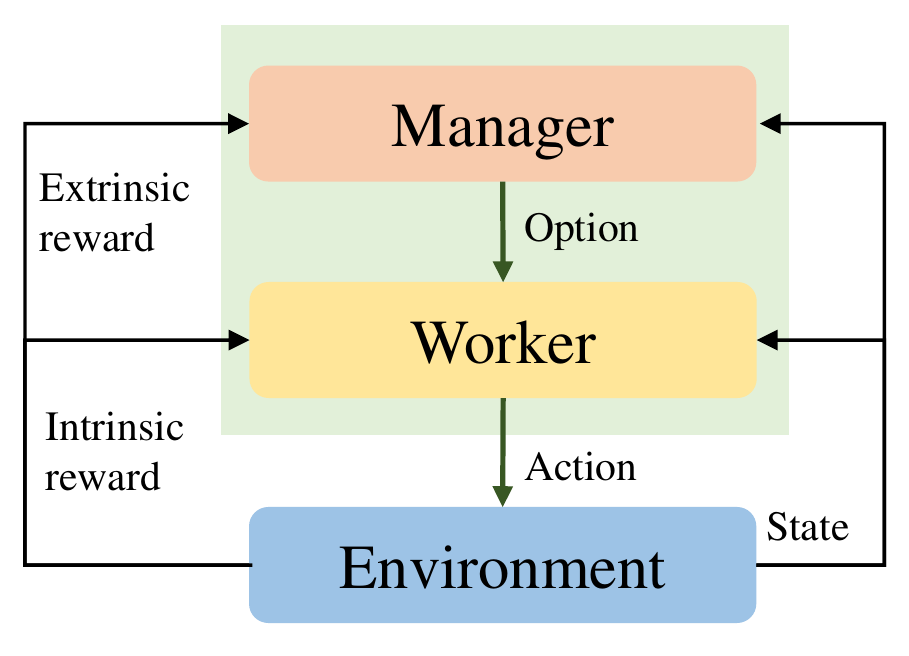}
 \caption{The feudal hierarchical reinforcement learning framework.}
 \label{fig:fhrl}
\end{figure}

Formally, we formulate the pair trading process as the feudal hierarchical reinforcement learning framework \cite{Pateria2021HierarchicalRL}. As shown in Fig.\ref{fig:fhrl}, a feudal hierarchical reinforcement learning framework consists of two controllers: a high-level controller called \textit{manager} and a low-level controller as \textit{worker}.
The manager is designed to set the option which aims to maximize the \textit{extrinsic} reward or the goal of the task. By selecting an option, the manager would trigger the worker which is guided by the \textit{intrinsic} reward. Different from the extrinsic reward as the overall target of the task, the intrinsic reward is a sub-goal of the manager given the selected option. Therefore, the decomposition allows the method to satisfy requirements at multiple levels to solve complex tasks that are infeasible for centralized reinforcement learning.

To this end, we design a high-level controller as the manager for pair trading, which aims to select two assets as a pair and maximize their trading performance via pair trading.
It is expected that the pair should possess the highest profit in the subsequent trading period among all possible combinations of assets.
Thus the extrinsic reward for the manager is the profit of the selected pair in the trading period.

However, to achieve the optimal trading profit of the selected pair, it is required to consider a different sub-task where the agent is supposed to perform a series of sequential trading decisions on the selected pair.
Since the target is different from the selection, we derive a low-level controller as the worker which only focuses on learning a flexible and profitable trading policy.
We adopt the profit of the selected pair in the formation period to guide the learning of the worker as the intrinsic reward.
After the worker is fully trained with the historical formation data, it is utilized to yield the trading performance of the selected pair with the unseen market data in the trading period, which is further taken as the extrinsic reward.

\subsection{Pair Selection with High-Level Controller}
For pair selection, we aim to select the optimal asset pair from all possible pairs of assets. It can be deemed as a contextual bandit~\cite{Langford2007TheEA} $M = (S^h, O, T^h, R^h, \Psi, Q^h)$ over options, where $S^h$ refers to the state space, $O$ is the option space, $T^h$ is the transitions among states, $R^h$ is the designed reward, $\Psi$ is the observation state which is generated from the current state $s^h \in S^h$ and the option referring to the high-level action $o \in O$ according to the probability distribution $Q^h(s^h, o)$.
Different from trading, the pair selection process is a one-step decision process that the agent would perform an option $o_0 \in O$ under the current state $s^h_0$, resulting in the transition from $s^h_0$ to $s^h_1$ with the probability $T(s^h_1|s^h_1, o_1)$.
After the option is selected, a low-level POMDP as the worker would be triggered to perform trading according to the selected option.

\subsubsection{Observation}
For the agent in the high-level contextual bandit, only limited information of the market state can be observed, i.e, the price features of the assets in history, which means the agent can only receive the observation $v_0 \in \Psi$ with probability as $Q(s^h_1, o_0)$.
The observation $v_0 \in \Psi$ is the price features for all assets $x \in \mathcal{X}$ associated with each time step $t \in T_f$ in the formation period, including the open price $p^o_{x,t}$, the close price $p^c_{x,t}$, and the volume $vol_{x,t}$.

\subsubsection{Option}
The option $o$ is a pair $(x_i, x_j)$ selected from all possible combinations of assets in $\mathcal{X}$.
When the low-level POMDP ended, the agent would select the next option according to the high-level contextual bandit.

\subsubsection{State}
Given the observation $v^h_0$ consisting of the open price $p^o_{x,t}$, the close price $p^c_{x,t}$, and the volume at each time point $vol_{x,t}$ for each asset $x \in \mathcal{X}$ in the formation period $t \in T_{\mathcal{F}}$,
we adopt the Bi-directional GRU (Bi-GRU)~\cite{hochreiter1997long} to capture the temporal correlations between historical price features.
Our method takes the previous hidden state $h_{t-1}$ as the hidden state of the forward GRU and the next state $h_{t}$ as the hidden state of the backward GRU.
Since the asset prices possess strong auto-correlation effects~\cite{10.1093/jjfinec/nbaa033}, it is fundamental to model the relationships from both history and future, which helps the method to capture salient information embedded in the asset price fluctuations.
Therefore, we represent our latent state $h_t$ as:
\begin{equation}
 \begin{split}
 \overrightarrow{h_t} = \text{GRU}(v^h_{0, t}, \overrightarrow{h_{t-1}}),\overleftarrow{h_t} = \text{GRU}(v^h_{0, t}, \overleftarrow{h_{t+1}}), h_t &= [\overrightarrow{h_t},\overleftarrow{h_t}]
 \end{split}
\end{equation}
where $h_t \in \mathbb{R}^{d_h}$ is the concatenation of the forward hidden state and backward hidden state, ${d_h}$ is the hidden dimension, and $v^h_{0, t} \in \mathbb{R}^{N \times 3}$ are the price features of all assets at the time step $t \in T_\mathcal{F}$ of the formation period.

As a matter of fact, Bi-GRU has the long-distance forgetting problem~\cite{bahdanau2014neural}, especially when there are thousands of time steps in the formation period.
We further introduce a temporal attention mechanism to dynamically select salient information from all historical time steps by:
\begin{align}
 \alpha_k &= \frac{
 exp(score(h_{T_\mathcal{F}}, h_k))
 }{
 \sum_{k'=0}^{T_\mathcal{F}-1}
 exp(score(h_{T_\mathcal{F}}, h_k'))
 },
	c_{T_\mathcal{F}} = \sum_k^{T_\mathcal{F}-1} \alpha_k h_k \\
	\hat{h}_{T_\mathcal{F}} &= LayerNorm(LeakyRelu(W_c[h_{T_\mathcal{F}}, c_{T_\mathcal{F}}]))
\end{align}
where $score(h_{T_\mathcal{F}}, h_k) = \frac{h_{T_\mathcal{F}} h_k}{\sqrt{d_h}}$ is the scaled dot-product attention score~\cite{NIPS2017_3f5ee243}.
We also adopt LeakyRelu and LayerNorm~\cite{https://doi.org/10.48550/arxiv.1607.06450} to stabilize the hidden state dynamics.
We adopt the final output $\hat{h}_{T_\mathcal{F}} \in \mathbb{R}^{N \times d_h}$ as the state $s^h_0 \in \mathbb{R}^{N \times d_h}$ of the high-level contextual bandit.
	
\subsubsection{Policy}
The stochastic policy for pair selection $\mu: \mathcal{S} \rightarrow \mathcal{O}$ refers to a probability distribution over options:
\begin{equation}
 o_0 \backsim \mu(o_0|s^h_0) = softmax(triu(s^h_0 {s^h_0}^T))
\end{equation}
where $triu$ is to extract and return the flattened upper triangular part of the given matrix.
	
\subsubsection{Reward}
 The reward of the high-level contextual bandit is the same as the target of the task, which is to maximize the profit of the trading period given the option $o_0$.
 However, realizing the optimal trading profit requires the method to learn a different sub-task.
 Therefore, we propose to utilize a low-level POMDP triggered by the selected option.
 It is first trained with the intrinsic reward in the formation period and then utilized to perform tradings to yield the trading profit in the trading period.
 Following previous RL-based trading methods, we also maximize the cumulative profit over the trading period with $T_\mathcal{T}$ time points:
\begin{equation}
 R^h = \prod_{t \in T_\mathcal{T}} (1 + R^h_t)
\end{equation}
where $R^h_t$ is the return of the low-level policy.
We would provide further details in the following subsections.
	
\subsection{Trading with Low-Level Controller}
When the high-level controller has selected a trading pair as the option, the low-level controller will perform trading based on the given trading pair as a series of trading actions in a subsequent trading period to achieve the trading profit.
Formally, we formulate the decision process of the trading as a Partially Observable Markov Decision Process (POMDP)~\cite{hausknecht2015deep} $M = (S^l, A, T^l, R^l, \Omega, Q^l)$, where $S^l$ refers to the state space, $A$ is the action space, $T^l$ is the transitions among states, $R^l$ is the designed reward, $\Omega$ is the partial observation state which is generated from the current state $s^l \in S^l$ and action $a \in A$ according to the probability distribution $Q^l(s^l, a)$.
At each time point, the agent would perform an action $a_t \in A$ under the current state $s^l_t$, resulting in the transition from $s^l_t$ to $s^l_{t+1}$ with the probability $T^l(s^l_{t+1}|s^l_t, a_t)$.
Similar to pair selection, the actual market states are partially observed and only the historical prices and volumes of assets, along with the historical account information of the agent such as actions, amounts of cash, and returns can be leveraged, while other information is ignored.
In detail, the agent can only receive the observation $v^l_{t+1} \in \Omega$ with probability as $Q(s^l_{t+1}, a_{t})$, which requires the agent to fully exploit the historical observations up to present time point.
	
\subsubsection{Observation}
The observation $v^l_t \in \Omega$ consists of two different feature sets, including: (1) the account features $v^a_t \in \Omega^a$ as previous action $a_{t-1}$, present cash $C_t$, present asset value $V_t$, and cumulative profit as the net value$N_t$; (2) the price features $o^p_t \in \Omega^p$ as the open price $p^o_{i,t}$, the close price $p^c_{i,t}$, and the volume $vol_{i,t}$ of for each asset $i \in \{X, Y\}$.
Following previous work~\cite{Lesmond2003TheIN}, we simplify the impact of tradings performed by our agent on the market state as a constant loss to each trading.
Therefore the action of our agent would not affect the state and price features of assets in our observation.
	
\subsubsection{Action}
The action in each time step is to perform a pair of contradictory trading actions on two assets respectively.
The action space $A = \{L, C, S\} = \{1, 0, -1\}$ consists of three discrete actions each of which involves two trading actions for two assets $\{X, Y\}$ respectively.
In detail, the \textit{L} action represents the long trading action which means to long asset $X$ and short asset $Y$ at the same time, the \textit{C} action for clear referring to clear two assets if longed or shorted any before, and the \textit{short} action for short which is to short asset $X$ and long asset $Y$.
Notice that for different asset pairs, the trading action at each time step could be assigned to other actions.
	
\subsubsection{State}
Different from the policy for pair selection,
the agent is required to estimate the latent market state $s^l_t$ according to the history $H_t = \{v^l_1, a_1, v^l_2, \cdots, a_{t-1}, v^l_t\}$.
Although the market state cannot be directly observed, the historical information embedded in $H_t$, especially the sequential dependencies can help the agent to generate better estimation.

Therefore, we also introduce Bi-GRU to encode the history.
The previous hidden state $h_{t-1}$ is deemed as the hidden state of the forward GRU and the next state $h_{t}$ as the hidden state of the backward GRU:
\begin{equation}
 \begin{split}
 \overrightarrow{h_t} = \text{GRU}(v^l_{0, t}, \overrightarrow{h_{t-1}}),\overleftarrow{h_t} = \text{GRU}(v^l_{0, t}, \overleftarrow{h_{t+1}}), h_t &= [\overrightarrow{h_t},\overleftarrow{h_t}]
 \end{split}
\end{equation}
where $h_t \in \mathbb{R}^{d_h}$ is also the concatenation of the forward hidden state and backward hidden state, ${d_h}$ is the hidden dimension, and $v^h_{0, t} \in \mathbb{R}^{2 \times M}$ are the input features of the selected two assets and $M$ is the feature dimension.
We transform discrete variables such as previous action $a_{t-1}$ in $v^a_t$ into continuous embeddings via an embedding layer $E_a \in \mathbb{R}^{3 \times d_a}$, where $d_a$ is the corresponding embedding size.

We also introduce a temporal attention mechanism:
\begin{align}
 \alpha_k &= \frac{
 exp(score(h_{t}, h_k))
 }{
 \sum_{k'=0}^{t-1}
 exp(score(h_t, h_k'))
 },
	c_{t} = \sum_k^{t-1} \alpha_k h_k \\
	\hat{h}_t &= LayerNorm(LeakyRelu(W_c[h_t, c_t]))
\end{align}
where $score(h_t, h_k) = \frac{h_t h_k}{\sqrt{d_h}}$ is the scaled dot-product attention score.
The output $\hat{h}_t \in \mathbb{R}^{d_h}$ as the $s^l_t \in \mathbb{R}^{d_h}$ of the low-level POMDP.
	
\subsubsection{Policy}
The stochastic policy for trading $\pi: \mathcal{S} \rightarrow \mathcal{A}$ yields a probability distribution over actions given the low-level state $s^l_t$ and the high-level option $o_0$:
\begin{equation}
 a_t \backsim \pi(a_t|s^l_t;o_0) = softmax(W_{\pi} s^l_t)
\end{equation}
	
\subsubsection{Reward}
 The intrinsic reward for the low-level controller is also the cumulative profit over a period with $T$ time points:
\begin{equation}
 R = \prod_{t \in T} (1 + R_t)
\end{equation}
where $R_t$ is the return of the agent given the action $a_t$:
\begin{equation}
 \begin{aligned}
 R_t & = a_{t-1} r_{X,t} - a_{t-1} r_{Y, t} - c|a_{t} - a_{t-1}| \\
 & = a_{t-1} (r_{X,t} - r_{Y,t}) - c|a_{t} - a_{t-1}|
 \end{aligned}
\end{equation}
Notice that the return of the agent is irrelevant to the market for hedging the return of two assets as $r_{X,t} - r_{Y,t}$.
To yield a positive return, it is required to select the optimal trading pair and precisely trading actions according to the historical performance of the trading pair.
For training, we use the formation period to guide the learning of the worker, and the trading period to yield a high-level extrinsic reward with the fully trained worker.
	
\subsection{Hierarchical Policy Learning}
For high-level policy updating, following the Advantage Actor-Critic method (A2C)~\cite{mnih2016asynchronous},
We update the policy and the value function every step as:
\begin{equation}
 \begin{aligned}
 & \nabla_{\theta_h^P} \log \mu(o_0|s^h_0; \theta_h^P) A(s^h_0; \theta_h^A) \\
 & \nabla_{\theta_h^A} \frac{1}{2} A^2(s^h_0; \theta_h^A)
 \end{aligned}
 \label{train_high_level}
\end{equation}
where $A(s^h_0; \theta_h^A)=r_1^h + \gamma V(s_1^h; \theta_h^{A-}) - V(s^h_0; \theta_h^A)$ is the estimation of the advantage function for the high-level controller and the option $o_0$ 
is sampled from the option distribution $\mu(o_0|s^h_0; \theta_h^P)$.
	
As for low-level policy updating, we apply A2C update similarly,
\begin{equation}
 \begin{aligned}
 & \nabla_{\theta_l^P} \log \pi(a_t|s^l_t;o_0, \theta_l^P) A(s^l_t; \theta_l^A) \\
 & \nabla_{\theta_l^A} \frac{1}{2} A^2(s^l_t; \theta_l^A)
 \end{aligned}
 \label{train_low_level}
\end{equation}
where $A(s^l_t; \theta_l^A)=r_{t+1}^l + \gamma V(s_{t+1}^l; \theta_l^{A-}) - V(s_t^l; \theta_l^A)$ is the estimation of the advantage function for the low-level controller and action $a_t$ is sampled from the action distribution $\pi(a_t|s^l_t;o_0, \theta_l^P)$.
	
For training, we adopt the same formation period data as the input to train both high-level and low-level policy, where the performance of the low-level policy during the trading period would be considered as the reward of the high-level policy.
As for evaluation and testing, we directly infer the option and corresponding actions without exploration.
	
\begin{algorithm}[H]
    \caption{Training}\label{training}
    \begin{algorithmic}[1]
        \REQUIRE $N$ assets $\mathcal{X}$, loop conditions $M, N$
        \ENSURE Model parameters $\theta_{h} = \{\theta_h^P, \theta_h^A\}, \theta_{l} = \{\theta_l^P, \theta_l^A\}$
        \STATE Initialize parameters $\theta_{h}, \theta_{l}$ for the high-level controller
        and low-level controller respectively
        \FOR{iteration=1, 2, 3, ..., M}
        \STATE Sample option $o_0$ from $\mu(o_0|s^h_0)$ 
        \STATE Select pair from $\mathcal{X}$ and initialize the trading environment
        \FOR{iteration=1, 2, 3, ..., N}
        \WHILE{not reach termination condition}
        \STATE Sample action $a_t$ from $\pi(a_t|s^l_t;o_0)$
        \STATE Execute action, then obtain the next state and intrinsic reward from the trading environment
        \STATE Update $\theta_l$ by Eq~(\ref{train_low_level})
        \ENDWHILE
        \ENDFOR
        \STATE Obtain extrinsic reward from pair selection environment
        \STATE Update $\theta_h$ by Eq~(\ref{train_high_level})
        \ENDFOR
    \end{algorithmic}
\end{algorithm}
	
\section{Experiments}
\subsection{Dataset}
Following previous methods~\cite{Krauss2017}, we build a dataset based on a pool of real stocks from S\&P 500\footnote{Tiingo. Tiingo stock market tools. https://api.tiingo.com/documentation/iex} for recent 21 years from 01/02/2000 to 12/31/2020.
We filter stocks that have missing data throughout the whole period, resulting in 150 stocks with 5,284 trading days.
To support the evaluation and development of pair trading, we introduce a new daily emerging stock market dataset (Chinese CSI 300 dataset) including 300 stocks and 5,088 time steps from the CSMAR database\footnote{www.gtarsc.com}.
Similar to previous work~\cite{Krauss2017}, we construct our stock dataset using a pool of stocks from the CSI 300 index for the last 21 years, from 01/02/2000 to 12/31/2020.
Instead of all stocks in the market, we select the stocks that used to belong to the major market index CSI 300 and filter out stocks that have missing price data over the period.
We compare our dataset and the U.S. stock market dataset S\&P 500 in Table \ref{tab:data-statistics}.
\begin{table}[htp]
\centering
\begin{tabular}{@{}ccccc@{}}
\toprule
Dataset & Market & Period & Assets & Time Steps \\ \midrule
S\&P 500 & U.S & 2000 - 2020 & 150 & 5284 \\\midrule
CSI 300 & China & 2000 - 2020 & 300 & 5088 \\ \bottomrule
\end{tabular}
\caption{The statistics of datasets.}
\label{tab:data-statistics}
\end{table}

For each trading day, we use the fundamental price features as the features of stocks, including open price, close price, and volume.
Additionally, we normalize price features such as open price and close price with logarithm.

Different from previous methods, we randomly split stocks into five non-overlapping sub-datasets, as shown in Appendix~\ref{app:stock-subset}.
For each subset with, we perform experiments of our method and baselines to evaluate their generality.
We use the first 90\% trading days as train data, the following 5\% as validation data, and the rest 5\% as test data.
For training, we further use the first 85\% trading days to train our methods to simultaneously select the optimal trading pair from  possible combinations and perform optimal trading actions based on the optimal trading pair in the rest of 5\% trading days.
The trained model is evaluated on the validation data to select the best hyperparameters based on which the performance of the model among the test data is reported.
We independently evaluate and report the performance of all methods on each subset, along with the mean and standard deviation over all subsets.
	
For our method and ablations, we use the RMSProp optimizer~\cite{tieleman2012rmsprop} and perform a bayesian parameter search~\cite{shahriari2015taking} for each subset to set the optimal hyper-parameters respectively. We implement our method based on Pytorch and stable-baselines, and conduct all our experiments on a server with 2 NVIDIA Tesla V100 GPUs.
	
\subsection{Baselines}
We compare our methods with the following baselines:
(1) \textbf{Pair selection methods}: they mainly focus on selecting the optimal asset pair which is expected to yield the best performance with threshold-based trading rules, such as \textbf{GGR}~\cite{gatev2006pairs} which uses average Euclidean distance to select pairs, \textbf{Cointegration}~\cite{vidyamurthy2004pairs} which adopts the augmented Engle-Granger two-step cointegration test to select the trading pair, and \textbf{Correlation}~\cite{elliott2005pairs} which selects two assets that have the highest correlation.
(2) \textbf{Trading methods}: they generally aim to train an agent to perform optimal trading actions with the asset pair which is generally selected using the augmented Engle-Granger two-step cointegration test, i.e, \textbf{Wang et al.} \cite{wang2021improving} that adopting the reinforcement learning to maximize the overall profit.
	
\subsection{Metrics}
As previous trading methods~\cite{wang2021improving}, we first evaluate our method along with baselines with their trading performance on the test data using 
(1) \textbf{Sharpe ratio (SR)} is the ratio of the profit to the risk~\cite{sharpe1994sharpe}, which is calculated as $(E(R_t)-R_f)/V(R_t)$, where $R_t$ is the daily return and $R_f$ is a risk-free daily return that is set to 0.000085 as previous methods.
(2) \textbf{Annualized return (AR)} is the expected profit of the agent when trading for a year. 
(3) \textbf{Maximum drawdown (MDD)} measures the risk as the maximum potential loss from a peak to a trough during the trading period. 
(4) \textbf{Annualized Volatility (AV)} measures the risk as the volatility of return over the course of a year.

We also employ fundamental measurements to measure the selected pair of all methods: average \textbf{Euclidean distance (ED)}~\cite{gatev2006pairs} which is the average euclidean distance of the historical price series of two assets.

\subsection{Main Results}
As shown in Table \ref{overall-result},
our method TRIALS achieves the best performance among all methods in all metrics and most stock subsets of both S\&P 500 and CSI 300.
\begin{table*}[htp]
\centering
\small
\begin{tabular}{@{}c||c|c|c|c|c||c|c@{}}
\toprule
Model &   & GGR & Cointegration & Correlation & Wang & TRIALS & TRIALS wo TR \\ \midrule
\multirow{5}{*}{S\&P 500} & SR$\Uparrow$ & \begin{tabular}[c]{@{}l@{}}-1.37\\ (0.79)\end{tabular} & \begin{tabular}[c]{@{}l@{}}-1.83\\ (0.27)\end{tabular} & \begin{tabular}[c]{@{}l@{}}-1.41\\ (0.21)\end{tabular} & \begin{tabular}[c]{@{}l@{}}1.18\\ (0.43)\end{tabular} & \textbf{\begin{tabular}[c]{@{}l@{}}1.84\\ (0.24)\end{tabular}} & \begin{tabular}[c]{@{}l@{}}0.07\\ (0.16)\end{tabular}  \\\cmidrule{2-8}
 & AR$\Uparrow$ & \begin{tabular}[c]{@{}l@{}}-0.15\\ (0.09)\end{tabular} & \begin{tabular}[c]{@{}l@{}}-0.36\\ (0.20)\end{tabular} & \begin{tabular}[c]{@{}l@{}}-0.14\\ (0.05)\end{tabular} & \begin{tabular}[c]{@{}l@{}}0.21\\ (0.11)\end{tabular} & \textbf{\begin{tabular}[c]{@{}l@{}}0.50\\ (0.14)\end{tabular}} & \begin{tabular}[c]{@{}l@{}}0.01\\ (0.20)\end{tabular} \\\cmidrule{2-8}
 & MDD$\Uparrow$ & \begin{tabular}[c]{@{}l@{}}-0.20\\ (0.08)\end{tabular} & \begin{tabular}[c]{@{}l@{}}-0.37\\ (0.20)\end{tabular} & \begin{tabular}[c]{@{}l@{}}-0.20\\ (0.04)\end{tabular} & \begin{tabular}[c]{@{}l@{}}-0.09\\ (0.05)\end{tabular} & \textbf{\begin{tabular}[c]{@{}l@{}}-0.09\\ (0.01)\end{tabular}} & \begin{tabular}[c]{@{}l@{}}-0.25\\ (0.07)\end{tabular} \\\cmidrule{2-8}
 & AV$\Downarrow$ & \begin{tabular}[c]{@{}l@{}}0.13\\ (0.03)\end{tabular} & \begin{tabular}[c]{@{}l@{}}0.27\\ (0.20)\end{tabular} & \textbf{\begin{tabular}[c]{@{}l@{}}0.12\\ (0.02)\end{tabular}} & \begin{tabular}[c]{@{}l@{}}0.16\\ (0.06)\end{tabular} & \begin{tabular}[c]{@{}l@{}}0.22\\ (0.04)\end{tabular} & \begin{tabular}[c]{@{}l@{}}0.22\\ (0.04)\end{tabular} \\\cmidrule{2-8}
 & ED$\Downarrow$ & \begin{tabular}[c]{@{}l@{}}0.014\\ (5e-3)\end{tabular} & \begin{tabular}[c]{@{}l@{}}0.021\\ (0.02)\end{tabular} & \textbf{\begin{tabular}[c]{@{}l@{}}0.007\\ (0.002)\end{tabular}} & \begin{tabular}[c]{@{}l@{}}0.021\\ (0.02)\end{tabular} & \begin{tabular}[c]{@{}l@{}}0.037\\ (0.01)\end{tabular} & \begin{tabular}[c]{@{}l@{}}0.01\\ (4e-3)\end{tabular}\\
 \midrule \midrule
 \multirow{5}{*}{CSI 300} & SR$\Uparrow$ & \begin{tabular}[c]{@{}l@{}}-1.19\\ (0.74)\end{tabular} & \begin{tabular}[c]{@{}l@{}}-1.50\\ (0.97)\end{tabular} & \begin{tabular}[c]{@{}l@{}}-1.37\\ (0.25)\end{tabular} & \begin{tabular}[c]{@{}l@{}}0.75\\ (0.68)\end{tabular} & \textbf{\begin{tabular}[c]{@{}l@{}}1.91\\ (0.88)\end{tabular}} & \begin{tabular}[c]{@{}l@{}}0.95\\ (0.88)\end{tabular} \\\cmidrule{2-8}
 & AR$\Uparrow$ & \begin{tabular}[c]{@{}l@{}}-0.17\\ (0.11)\end{tabular} & \begin{tabular}[c]{@{}l@{}}-0.25\\ (0.17)\end{tabular} & \begin{tabular}[c]{@{}l@{}}-0.21\\ (0.07)\end{tabular} & \begin{tabular}[c]{@{}l@{}}0.24\\ (0.23)\end{tabular} & \textbf{\begin{tabular}[c]{@{}l@{}}0.68\\ (0.51)\end{tabular}} & \begin{tabular}[c]{@{}l@{}}0.13\\ (0.12)\end{tabular}\\\cmidrule{2-8}
 & MDD$\Uparrow$ & \begin{tabular}[c]{@{}l@{}}-0.29\\ (0.06)\end{tabular} & \begin{tabular}[c]{@{}l@{}}-0.29\\ (0.13)\end{tabular} & \begin{tabular}[c]{@{}l@{}}-0.25\\ (0.06)\end{tabular} & \begin{tabular}[c]{@{}l@{}}-0.18\\ (0.09)\end{tabular} & \begin{tabular}[c]{@{}l@{}}-0.14\\ (0.07)\end{tabular} & \textbf{\begin{tabular}[c]{@{}l@{}}-0.12\\ (0.09)\end{tabular}}\\\cmidrule{2-8}
 & AV$\Downarrow$ & \begin{tabular}[c]{@{}l@{}}0.18\\ (0.03)\end{tabular} & \begin{tabular}[c]{@{}l@{}}0.19\\ (0.03)\end{tabular} & \textbf{\begin{tabular}[c]{@{}l@{}}0.17\\ (0.05)\end{tabular}} & \begin{tabular}[c]{@{}l@{}}0.25\\ (0.07)\end{tabular} & \begin{tabular}[c]{@{}l@{}}0.26\\ (0.09)\end{tabular} & \begin{tabular}[c]{@{}l@{}}0.17\\ (0.07)\end{tabular} \\\cmidrule{2-8}
 & ED$\Downarrow$ & \textbf{\begin{tabular}[c]{@{}l@{}}0.013\\ (6e-3)\end{tabular}} & \begin{tabular}[c]{@{}l@{}}0.017\\ (8e-3)\end{tabular} & \begin{tabular}[c]{@{}l@{}}0.015\\ (8e-3)\end{tabular} & \begin{tabular}[c]{@{}l@{}}0.017\\ (8e-3)\end{tabular} & \begin{tabular}[c]{@{}l@{}}0.046\\ (0.02)\end{tabular} & \begin{tabular}[c]{@{}l@{}}0.02\\ (8e-3)\end{tabular} \\
 \bottomrule
\end{tabular}

\caption{Mean(Standard Deviation) of all metrics on S\&P 500 and CSI 300.}
\label{overall-result}
\end{table*}
The detailed performance of each subset in two datasets is presented in Appendix~\ref{app:detail-performance}.
It demonstrates the effectiveness of our method for simultaneously learning the pair selection and trading with a unified hierarchical reinforcement learning framework.
In detail, TRIALS has the highest average SR and AR, which indicates that our trained trading agent can yield a remarkable profit with controlled risks based on the selected pair of our method.
Our method presents a consistently high performance in two datasets, which clearly shows the two tasks in our unified framework are complementary since the pair selection task requires the trading performance of assets while the trading task depends on the selected pair.
This is further proved when TRIALS also yields the lowest average MDD in S\&P 500 and a relatively low MDD in CSI 300.
Since AV indicates both the fluctuations during rises and falls, our method presents a relatively high average AV in both datasets.

In contrast, previous pair selection methods such as GGR, Cointegration, and Correlation, underperform our method.
The average SR of GGR is -1.37 and -1.19, and its AR in both two datasets are negative.
Cointegration has an even worse average SR of -1.83 and -1.50, and also negative AR in both datasets.
Similarly, correlation shows an average SR as -1.41 and -1.37 along with a negative SR.
It clearly shows that pair selection methods based on pre-defined statistical tests or fundamental measurements would fail to select the optimal trading pair
without considering the trading performance of the asset pairs.
Moreover, the statistical tests and fundamental measurements adopted in their methods cannot measure the profitability of the asset pair even with the test data.
For example, our method has a higher ED compared with existing methods, despite the fact that our method yields a significant profit in both datasets.

As for trading methods such as Wang et al. which adopt reinforcement learning to train a flexible agent, it shows a better performance than pair trading methods such as Cointegration with the same selected pair.
However, it is limited by the selected pair based on the cointegration test which is shown to be ineffective in capturing the profitability of the asset pair, resulting in a lower SR compared with our method.
	
\subsection{Ablation Study}
To evaluate the contributions of two tasks in our unified framework, we further propose an ablation of our proposed method to compare with our method, as shown in Table~\ref{overall-result}, which is \textbf{TRIALS wo TR} that adopts a fixed trading agent with predefined thresholds after our RL based pair selection.

Compared with TRIALS wo TR, our method which jointly optimizes the two tasks presents the best performance with the highest average SR and lowest average ED.
In contrast, TRIALS wo TR is misguided by the trading performance of fixed trading rules which strongly relies on the wrong estimation of the mean and standard deviation of the price spread as the historical mean and standard deviation, resulting in a worse result in comparison to our method.

However, TRIALS wo TR still outperforms existing parameter-free methods, which proves the importance of dynamically learning the measurement of the future profitability according to their trading performance in pair selection.

\begin{figure}[ht]
    \centering
    \includegraphics[width=\columnwidth]{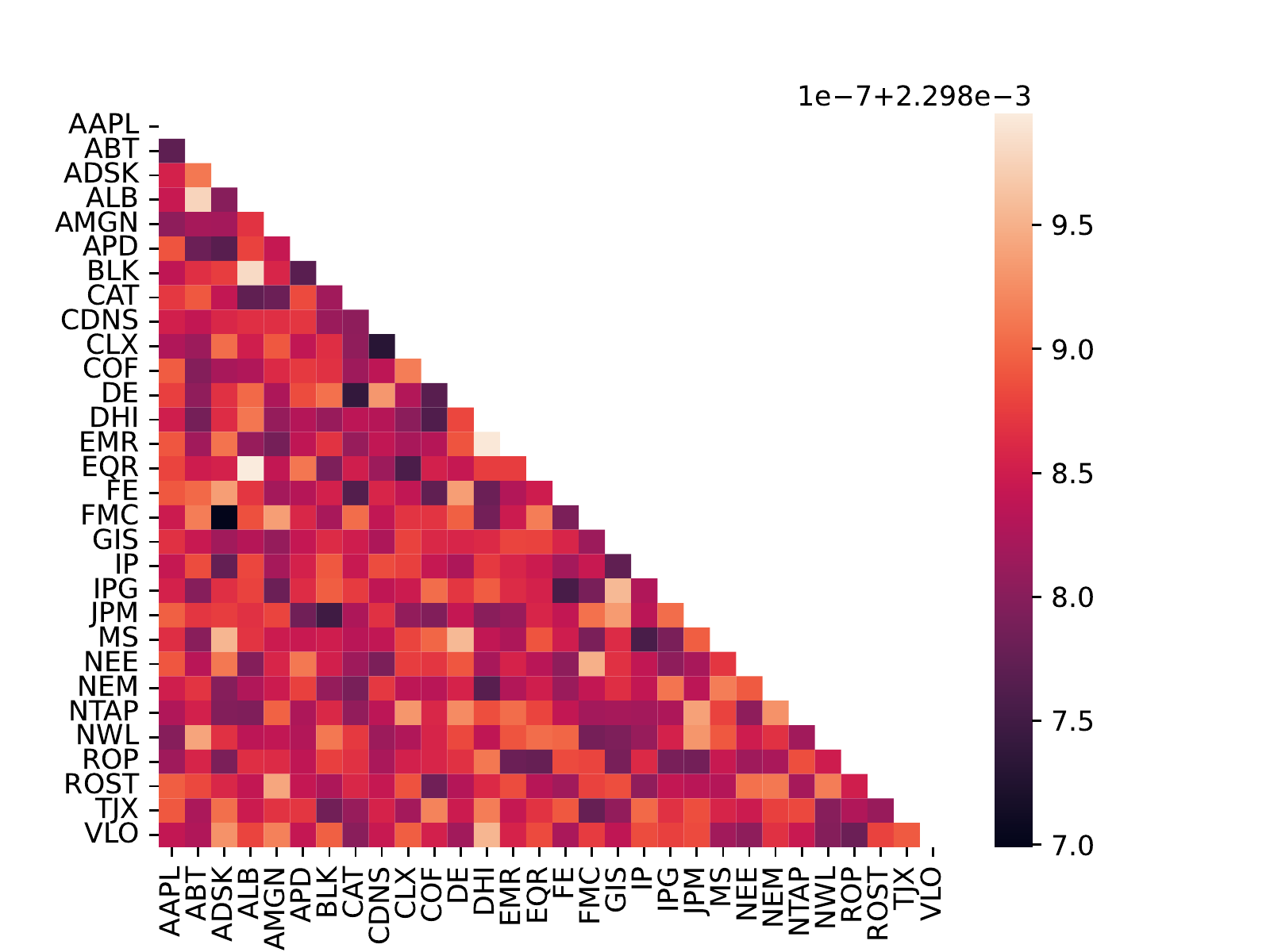}
    \caption{The pair selection probabilities of TRIALS.}
    \label{fig:pair-prob}
\end{figure}

\begin{figure}[ht]
\centering
\includegraphics[width=\columnwidth]{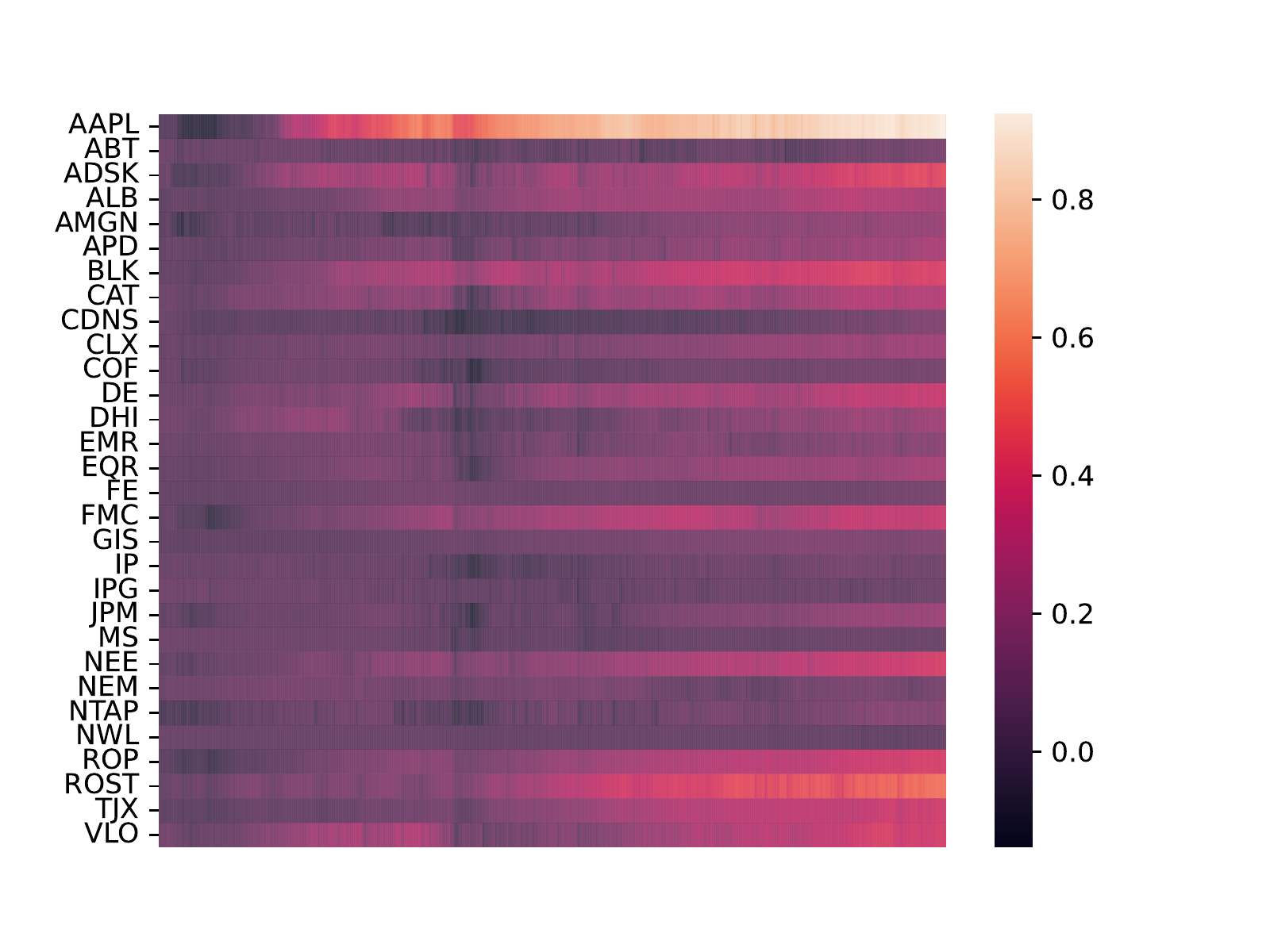}
\caption{The visualization of temporal attentions}
\label{fig:attentions}
\end{figure}

We also display the visualizations of the learned pair selection probability of our method in Fig.\ref{fig:pair-prob}.
It clearly shows that our method can precisely capture the complex connections between asset pairs.
For example, EQR engages in the real estate investment and ABT engages in chemicals.
Although there are no direct connections, our method finds that they are strongly and consistently correlated, which indicates a more complex multi-hop relationship between these two stocks such as industry spillover.
Besides, we display the temporal attention of 
our method in Fig.\ref{fig:attentions}.
As shown in Fig.\ref{fig:attentions}, our method can fully exploit the temporal information by temporal attention, which means, for AAPL, we would focus more on the latest features.
\subsection{Case Study}
To further verify the profitability of the selected pair and learned trading agent of our method,
we show the detailed trading actions, positions, and profit during the trading period of the selected pair by TRIALS, TRIALS wo TR, GGR, and Wang et al. in Set 2, as shown in Fig.\ref{fig:trading-case}.
The larger version of these figures is presented in Appendix~\ref{app:result-presentation}.
\begin{figure*}[htb]
\centering
\begin{subfigure}[b]{\columnwidth}
    \centering
    \includegraphics[width=\columnwidth]{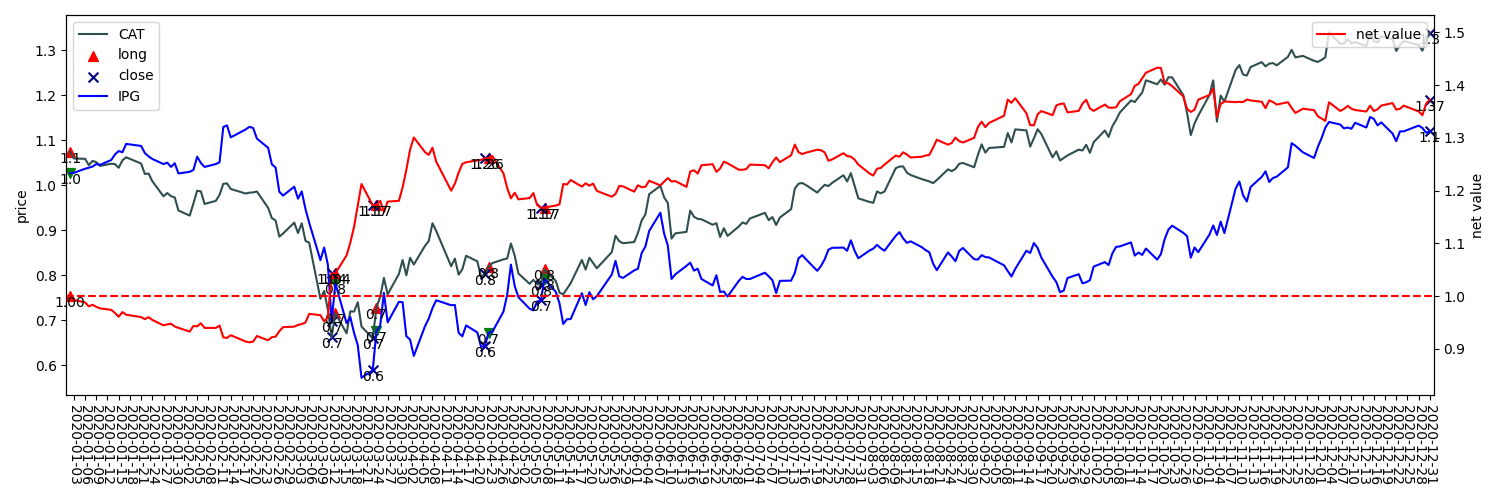}
    \caption{The trading detail of TRIALS.}
    \label{fig:trading-case:TRIALS}
\end{subfigure}
\hfill
\begin{subfigure}[b]{\columnwidth}
    \centering
    \includegraphics[width=\columnwidth]{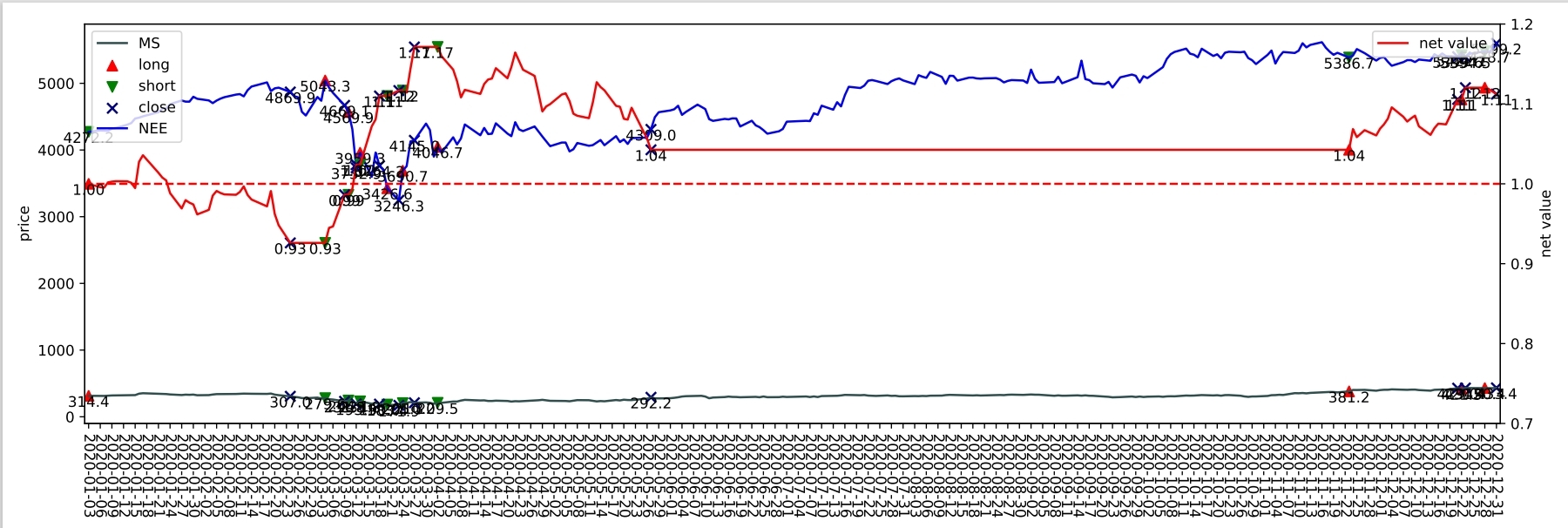}
    \caption{The trading detail of TRIALS wo TR.}
    \label{fig:trading-case:L2P}
\end{subfigure}
\hfill
\begin{subfigure}[b]{\columnwidth}
    \centering
    \includegraphics[width=\columnwidth]{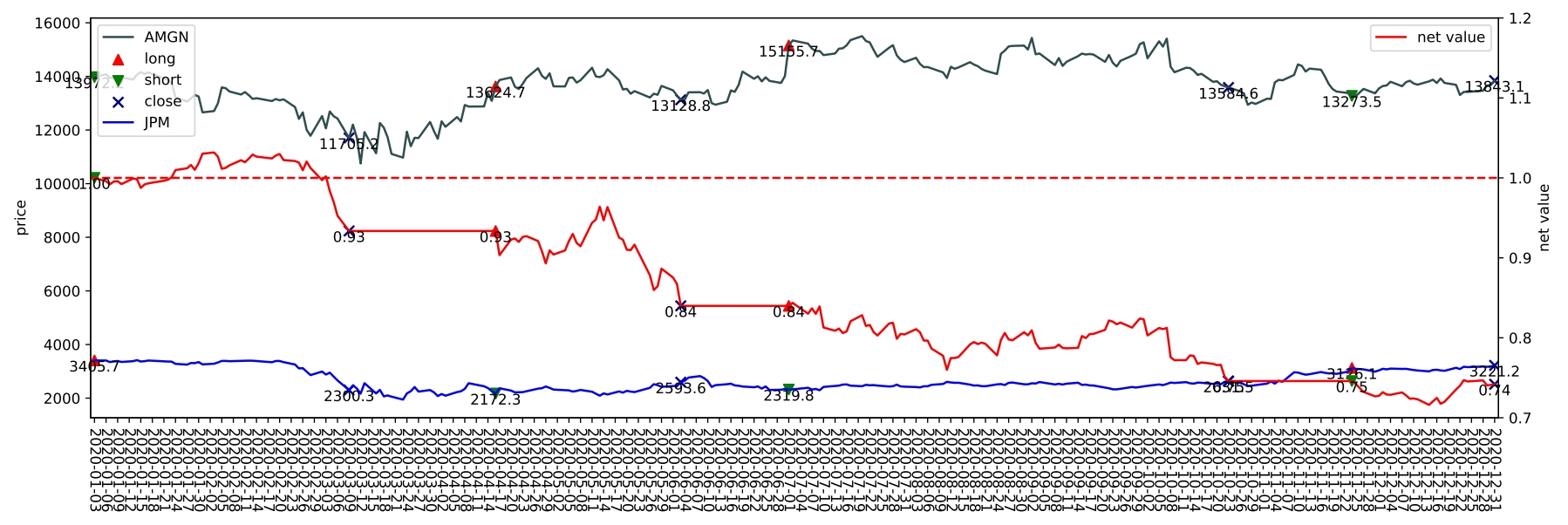}
    \caption{The trading detail of GGR.}
    \label{fig:trading-case:ggr}
\end{subfigure}
\hfill
\begin{subfigure}[b]{\columnwidth}
    \centering
    \includegraphics[width=\columnwidth]{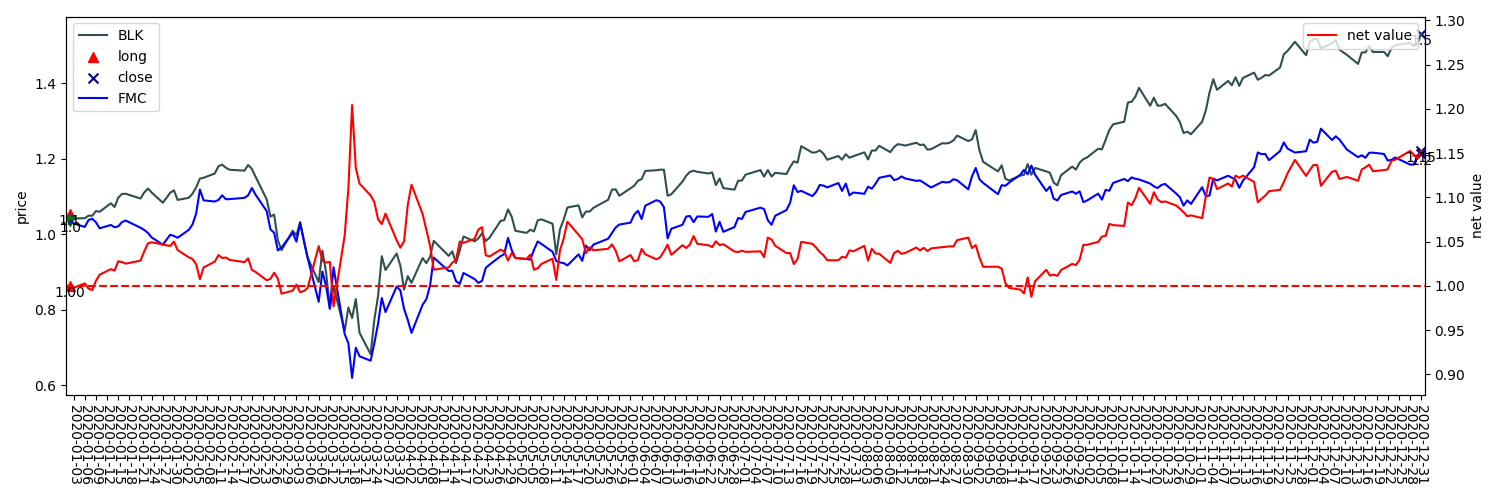}
    \caption{The trading detail of Wang et al.}
    \label{fig:trading-case:wang}
\end{subfigure}
\hfill
\caption{The trading details}
\label{fig:trading-case}
\end{figure*}
Since GGR and Wang et al. both ignore the trading performance of assets, there are irrelevant movements of the prices of the selected pair such as NEE and MS, resulting in wrong tradings with great loss.
As for GGR and TRIALS w/o TR, they show irrational trading decisions due to the wrongly-estimated thresholds for trading, also leading to poor trading performance.

Compared with them, our method that jointly learns to select optimal pair and trade can simultaneously consider the information of all assets and dynamically learns the measurement according to the optimal trading performance based on a flexible agent.
It allows our method to select the profitable pair CAT and IPG, which shows multiple trading opportunities in the trading period which are precisely captured by our trained trading method.
Besides, our method can observe multiple asset pairs, which forces the worker to capture the consistent pattern for pair trading instead of overfitting to only one selected asset pair.
Thus the worker in our method can precisely capture the trading opportunities and yield significant profit.

In contrast, although TRIALS w/o TR also learns to dynamically select asset pairs according to their trading performance, the fixed-threshold-based trading method can only provide biased information, resulting in less profitable pair selection.

\section{Conclusion}
In this paper, we proposed a novel paradigm for automatic pair trading that unifies the two sub-tasks: pair selection and trading.
Based on it, we designed a feudal hierarchical reinforcement learning method consisting of a high-level manager for selection and a low-level worker for trading.
The manager focused on selecting a pair as the option from all possible combinations of assets to maximize its trading performance, while the worker was to achieve the option set by the manager and yield the trading performance of the selected pair after training on historical data.
Experimental results on the real-world stock data prove that the two steps in pair trading are closely related and complementary, which our method can fully exploit and jointly optimize to generate a significant improvement compared with existing pair selection and trading methods.
In the future, we would further integrate more representation methods for learning the representations of assets and consider other information such as natural language texts and macroeconomic variables.

\bibliographystyle{ACM-Reference-Format}
\bibliography{trials}

\appendix
\section{Stock Subset}
\label{app:stock-subset}
In this section, we present the stocks of all 5 randomly split subgroups in the U.S. and Chinese stock markets in Table \ref{tab:stock-subset-us} and Table \ref{tab:stock-subset-cn} respectively.
\begin{table}[htp]
	\centering
	\small
	\begin{tabular}{@{} c @{} || l  @{}}
		\toprule
		Set & Stocks \\\midrule\midrule
		1 & \makecell[l]{AMAT, AXP, BA, BAX, EA, EBAY, ED, EOG,\\
        GLW, IBM, IRM, LMT, MAS, MCO, MMM, MOS,\\
        NUE, PFE, PG, PPL, QCOM, RTX, SLB, SPG,\\
        SWKS, TGT, TXT, UNH, USB, WY} \\ \midrule
		2 & \makecell[l]{AAPL, ABT, ADSK, ALB, AMGN, APD, BLK, CAT,\\
        CDNS, CLX, COF, DE, DHI, EMR, EQR, FE,\\
        FMC, GIS, IP, IPG, JPM, MS, NEE, NEM,\\
        NTAP, NWL, ROP, ROST, TJX, VLO} \\ \midrule
		3 & \makecell[l]{ADBE, AES, AVY, BSX, C, CAH, CCL, CL,\\CMI, CTSH, DOV, DUK, EXC, F, GE, HSY,\\
        KO, KR, LUV, MRO,MSFT, NKE, PEAK, PLD,\\
        PNC, SCHW, SYY, UPS, VFC, YUM} \\ \midrule
		4 & \makecell[l]{A, ADM, ALL, ATVI, AZO, BMY, COST, CSCO,\\
        CVX, FCX, FDX, GS, HAL, HD, INTC, K,\\
        KIM, LEN, LOW, MCD,MMC, MRK, MSI, NVDA,\\
        PHM, STT, T, WMB, XOM, XRAY} \\ \midrule
		5 & \makecell[l]{AMZN, AON, APA, BAC, BBY, BEN, BK, CMCSA,\\
        CPRT, CVS,DHR, EIX, ETN, FAST, HON, HUM,\\
        MCK, MO, MTB, NLOK, PCAR, PGR, SBUX, TER,\\
        TRV, UNP, VZ, WFC, WHR, WMT} \\ \midrule
		
	\end{tabular}
		\caption{The stocks of subsets for S\&P 500.}
	\label{tab:stock-subset-us}
\end{table}
\begin{table}[htp]
	\centering
	\small
	\begin{tabular}{@{} c @{} || c  @{}}
		\toprule
		Set & Stocks \\\midrule\midrule
		1 & \makecell[l]{000012,000016,000022,000024,000027,000029,
        000046,000059,\\000068,000088,000422,000518,
        000520,000539,000541,000559,\\000568,000598,
        000625,000627,000671,000698,000708,000712,\\
        000717,000729,000776,000825,000831,000878,
        000916,000920,\\000933,600061,600108,600109,
        600118,600125,600138,600151,\\600183,600228,
        600239,600602,600611,600641,600642,600688,\\
        600710,600717,600744,600770,600795,600811,
        600820,600832,\\600875,600884,600886,600893} \\ \midrule
		2 & \makecell[l]{000009,000089,000401,000402,000420,000425,000503,000507,\\
        000536,000538,000553,000623,000709,000758,000806,000900,\\
        000912,000927,000937,000938,600005,600009,600057,600058,\\
        600066,600078,600111,600115,600123,600150,600153,600157,\\
        600170,600177,600190,600196,600266,600600,600601,600606,\\
        600639,600643,600649,600662,600724,600726,600780,600783,\\
        600797,600804,600809,600812,600835,600838,600863,600879,\\600880,600881,600887,600894} \\ \midrule
		3 & \makecell[l]{000031,000069,000423,000532,000562,000571,000599,000601,\\
        000630,000631,000651,000661,000666,000690,000738,000750,\\
        000767,000768,000778,000780,000793,000823,000828,000898,\\
        000921,000930,000932,600072,600079,600085,600088,600091,\\
        600096,600100,600104,600117,600135,600169,600171,600198,\\
        600200,600210,600216,600608,600630,600635,600657,600663,\\
        600704,600707,600718,600737,600739,600747,600823,600839,\\
        600866,600868,600873,600874} \\ \midrule
		4 & \makecell[l]{000021,000036,000061,000063,000066,000408,000413,000498,\\
        000528,000533,000550,000596,000612,000667,000682,000686,\\
        000703,000707,000718,000728,000735,000737,000786,000883,\\
        000886,000897,000939,000951,600006,600062,600074,600098,\\
        600103,600110,600121,600126,600176,600215,600219,600220,\\
        600221,600621,600633,600648,600654,600655,600660,600675,\\
        600705,600748,600757,600761,600779,600808,600816,600827,\\
        600837,600854,600867,600895} \\ \midrule
		5 & \makecell[l]{000001,000002,000039,000060,000400,000415,000429,000543,\\
        000573,000581,000607,000629,000636,000652,000656,000659,\\
        000680,000685,000727,000733,000783,000792,000800,000807,\\
        000822,000826,000839,000858,000876,000895,000917,000949,\\
        000959,600000,600007,600060,600068,600073,600089,600132,\\
        600161,600188,600207,600208,600637,600638,600652,600653,\\
        600664,600674,600690,600694,600703,600733,600741,600760,\\
        600790,600805,600851,600871} \\ \midrule
		
	\end{tabular}
		\caption{The stocks of subsets for CSI 300.}
	\label{tab:stock-subset-cn}
\end{table}
\section{Result presentation}
\label{app:result-presentation}
\begin{figure*}[htb]
\centering
\begin{subfigure}[b]{.8\textwidth}
\centering
\includegraphics[width=\columnwidth]{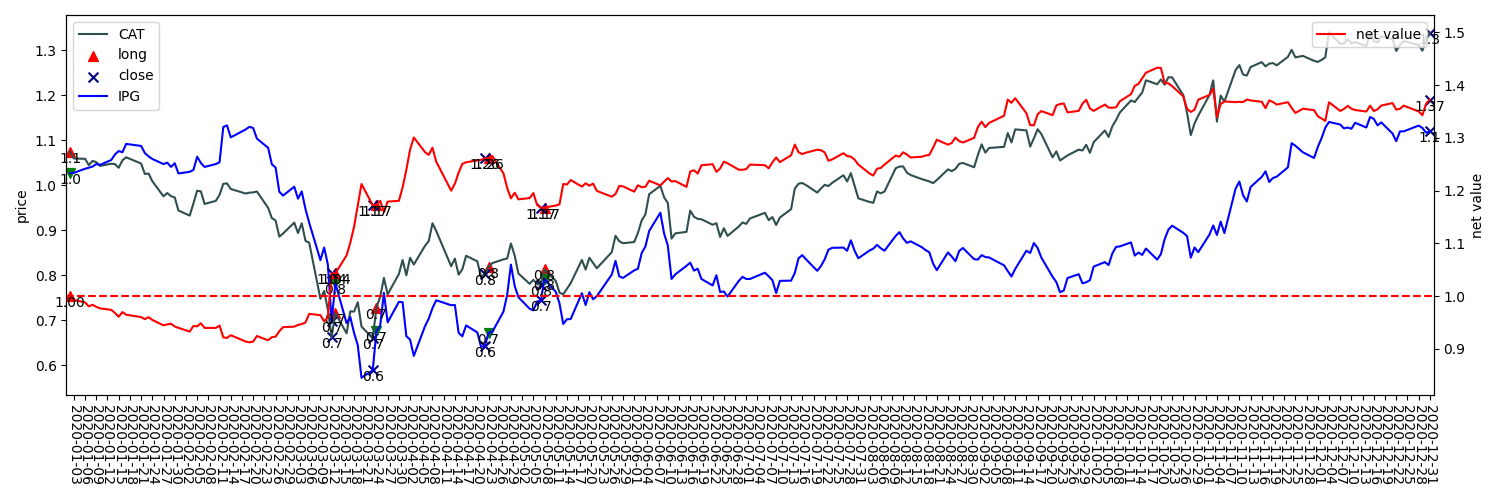}
\caption{The trading detail of TRIALS.}
\label{fig:detail-trading-case:TRIALS}
\end{subfigure}
\hfill
\begin{subfigure}[b]{.8\textwidth}
\centering
\includegraphics[width=\columnwidth]{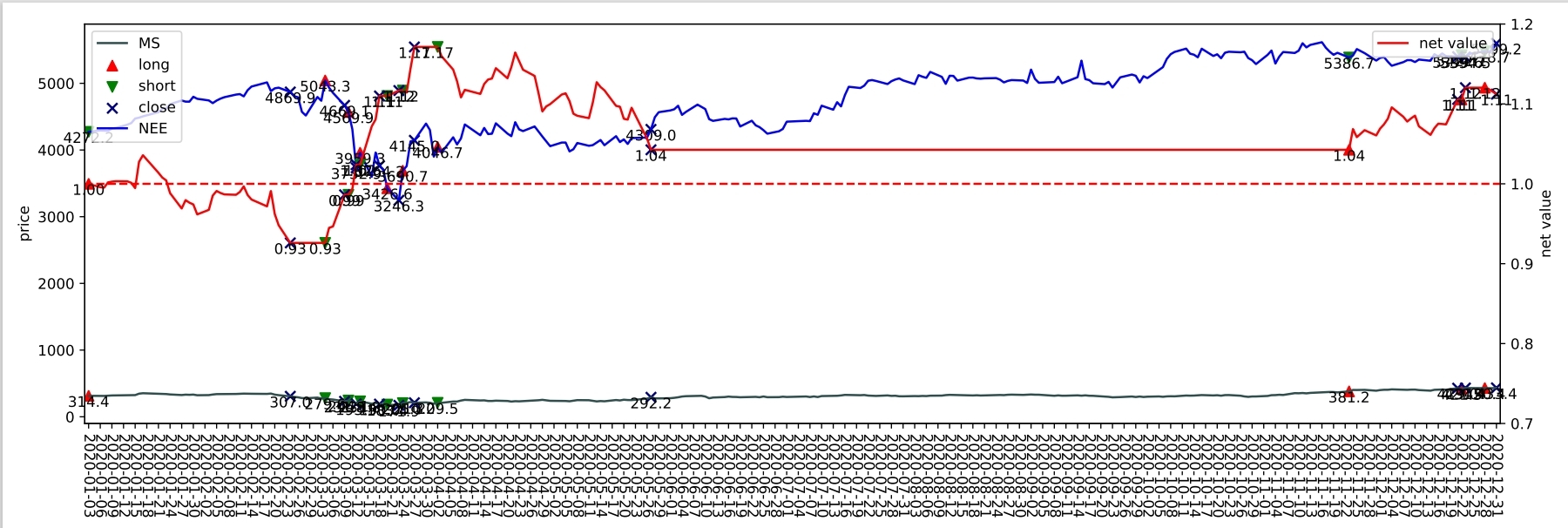}
\caption{The trading detail of TRIALS w/o TR.}
\label{fig:detail-trading-case:L2P}
\end{subfigure}
\hfill
\begin{subfigure}[b]{.8\textwidth}
\centering
\includegraphics[width=\columnwidth]{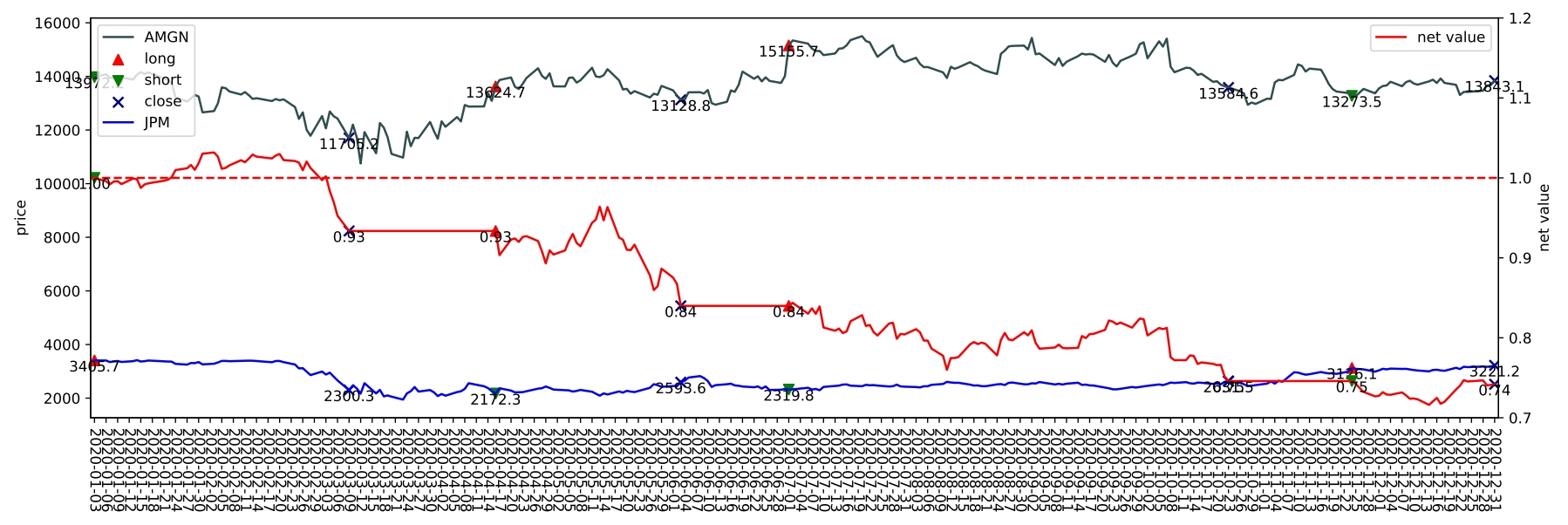}
\caption{The trading detail of GGR.}
\label{fig:detail-trading-case:ggr}
\end{subfigure}
\hfill
\begin{subfigure}[b]{.8\textwidth}
\centering
\includegraphics[width=\columnwidth]{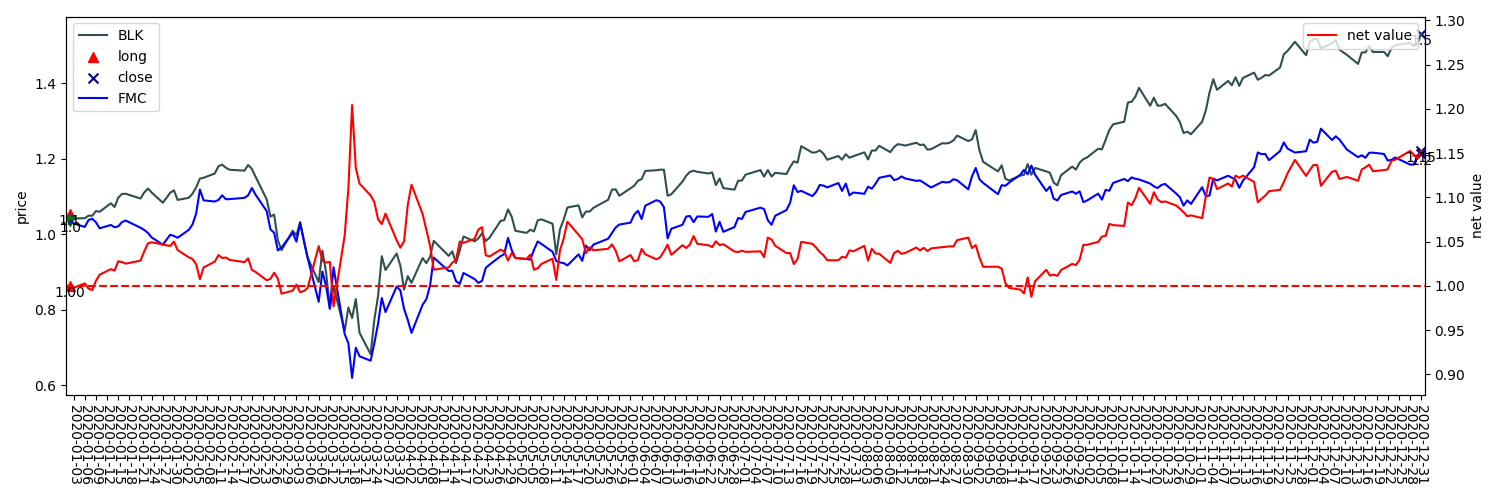}
\caption{The trading detail of Wang et al.}
\label{fig:detail-trading-case:wang}
\end{subfigure}
\hfill
\caption{The trading details}
\label{fig:detail-trading-case}
\end{figure*}

We present a larger version of the detailed trading actions, positions, and profit during the trading period of the selected pair by BanditPair and GGR in Set 2 of S\&P 500, as shown in Fig.\ref{fig:detail-trading-case}.

\section{Detail Performance}
\label{app:detail-performance}
As shown in Table \ref{overall-result-us} and Table \ref{overall-result-cn}, we report in this section the performance of each approach over all 5 subgroups of the U.S. and Chinese stock markets.

\begin{table*}[htp]
\centering
\small
\begin{tabular}{@{}l||c|c|c|c|c||c|c|@{}}
\toprule
Model &  & GGR & Cointegration & Correlation & Wang & TRIALS & TRIALS wo TR\\ \midrule
\multirow{5}{*}{Set 1} & SR$\Uparrow$ & -0.39 & -1.56 & -1.26 & 1.32 & \textbf{1.57} & -0.08  \\ \cmidrule{2-8}
& AR$\Uparrow$ & -0.03 & -0.39 & -0.14 & 0.32 & \textbf{0.51} & -0.04  \\\cmidrule{2-8}
& MDD$\Uparrow$ & -0.15 & -0.41 & -0.23 & -0.11 & \textbf{-0.09} & -0.32  \\\cmidrule{2-8}
 & AV$\Downarrow$ & \textbf{0.12} & 0.30 & 0.13 & 0.21 & 0.27 & 0.28  \\\cmidrule{2-8}
& ED$\Downarrow$ & \textbf{0.008} & 0.035 & 0.010 & 0.036 & 0.027 & 0.011  \\ \midrule \midrule
\multirow{5}{*}{Set 2} & SR$\Uparrow$ & -2.01 & -1.84 &-1.58 & 0.73 & \textbf{2.10} & -0.08 \\\cmidrule{2-8}
 & AR$\Uparrow$ & -0.29 & -0.20 & -0.19 & 0.14 & \textbf{0.64} & -0.01\\\cmidrule{2-8}
 & MDD$\Uparrow$ & -0.33 & -0.25 & -0.22 & \textbf{-0.08} & \-0.11 & -0.21  \\\cmidrule{2-8}
 & AV$\Downarrow$ & 0.17 & \textbf{0.13} & 0.14 & 0.17 & 0.24 & 0.19 \\\cmidrule{2-8}
 & ED$\Downarrow$ & 0.015 & 0.008 & \textbf{0.006} & 0.008 & 0.037 & 0.010 \\\midrule \midrule
\multirow{5}{*}{Set 3} & SR$\Uparrow$ & -1.49 & -1.74 & -1.30 & 1.70 & \textbf{1.86} & 0.02  \\\cmidrule{2-8}
 & AR$\Uparrow$ & -0.20 & -0.74 & -0.10 & \textbf{0.36} & 0.34 & 0.10  \\\cmidrule{2-8}
 & MDD$\Uparrow$ & -0.24 & -0.74 & -0.13 & -0.10 & \textbf{-0.08} & -0.20 \\\cmidrule{2-8}
 & AV$\Downarrow$ & 0.16 & 0.65 & \textbf{0.09} & 0.18 & 0.15 & 0.20  \\\cmidrule{2-8}
 & ED$\Downarrow$ & 0.022 & 0.044 & \textbf{0.005} & 0.005 & 0.029 & 0.008 \\\midrule \midrule
\multirow{5}{*}{Set 4} & SR$\Uparrow$ & -0.55 & -1.66 & -1.16 & 1.51 & \textbf{2.10} & 0.15 \\\cmidrule{2-8}
 & AR$\Uparrow$ & -0.06 & -0.18 & -0.08 & 0.10 & \textbf{0.67} & 0.03 \\\cmidrule{2-8}
 & MDD$\Uparrow$ & -0.10 & -0.20 & -0.16 & \textbf{0} & -0.10 & -0.33 \\\cmidrule{2-8}
 & AV$\Downarrow$ & 0.14 & 0.13 & 0.09 & \textbf{0.05} & 0.25 & 0.24 \\\cmidrule{2-8}
 & ED$\Downarrow$ & 0.016 & 0.006 & \textbf{0.006} & 0.006 & 0.060 & 0.302 \\\midrule \midrule
\multirow{5}{*}{Set 5} & SR$\Uparrow$ & -2.42 & -2.35 & -1.73 & 0.61 & \textbf{1.56} & 0.35 \\\cmidrule{2-8}
 & AR$\Uparrow$ & -0.16 & -0.27 & -0.20 & 0.13 & \textbf{0.34} & 0.07 \\\cmidrule{2-8}
 & MDD$\Uparrow$ & -0.19 & -0.27 & -0.25 & -0.14 & \textbf{-0.09} & -0.18 \\\cmidrule{2-8}
 & AV$\Downarrow$ & \textbf{0.08} & 0.13 & 0.13 & 0.19 & 0.18 & 0.18  \\\cmidrule{2-8}
 & ED$\Downarrow$ & 0.010 & 0.009 & \textbf{0.007} & 0.007 & 0.033 & 0.009 \\\midrule \midrule
\multirow{5}{*}{\begin{tabular}[c]{@{}l@{}}Mean\\ (Std)\end{tabular}} & SR$\Uparrow$ & \begin{tabular}[c]{@{}l@{}}-1.37\\ (0.79)\end{tabular} & \begin{tabular}[c]{@{}l@{}}-1.83\\ (0.27)\end{tabular} & \begin{tabular}[c]{@{}l@{}}-1.41\\ (0.21)\end{tabular} & \begin{tabular}[c]{@{}l@{}}1.18\\ (0.43)\end{tabular} & \textbf{\begin{tabular}[c]{@{}l@{}}1.84\\ (0.24)\end{tabular}} & \begin{tabular}[c]{@{}l@{}}0.07\\ (0.16)\end{tabular}  \\\cmidrule{2-8}
 & AR$\Uparrow$ & \begin{tabular}[c]{@{}l@{}}-0.15\\ (0.09)\end{tabular} & \begin{tabular}[c]{@{}l@{}}-0.36\\ (0.20)\end{tabular} & \begin{tabular}[c]{@{}l@{}}-0.14\\ (0.05)\end{tabular} & \begin{tabular}[c]{@{}l@{}}0.21\\ (0.11)\end{tabular} & \textbf{\begin{tabular}[c]{@{}l@{}}0.50\\ (0.14)\end{tabular}} & \begin{tabular}[c]{@{}l@{}}0.01\\ (0.20)\end{tabular} \\\cmidrule{2-8}
 & MDD$\Uparrow$ & \begin{tabular}[c]{@{}l@{}}-0.20\\ (0.08)\end{tabular} & \begin{tabular}[c]{@{}l@{}}-0.37\\ (0.20)\end{tabular} & \begin{tabular}[c]{@{}l@{}}-0.20\\ (0.04)\end{tabular} & \begin{tabular}[c]{@{}l@{}}-0.09\\ (0.05)\end{tabular} & \textbf{\begin{tabular}[c]{@{}l@{}}-0.09\\ (0.01)\end{tabular}} & \begin{tabular}[c]{@{}l@{}}-0.25\\ (0.07)\end{tabular} \\\cmidrule{2-8}
 & AV$\Downarrow$ & \begin{tabular}[c]{@{}l@{}}0.13\\ (0.03)\end{tabular} & \begin{tabular}[c]{@{}l@{}}0.27\\ (0.20)\end{tabular} & \textbf{\begin{tabular}[c]{@{}l@{}}0.12\\ (0.02)\end{tabular}} & \begin{tabular}[c]{@{}l@{}}0.16\\ (0.06)\end{tabular} & \begin{tabular}[c]{@{}l@{}}0.22\\ (0.04)\end{tabular} & \begin{tabular}[c]{@{}l@{}}0.22\\ (0.04)\end{tabular} \\\cmidrule{2-8}
 & ED$\Downarrow$ & \begin{tabular}[c]{@{}l@{}}0.014\\ (5e-3)\end{tabular} & \begin{tabular}[c]{@{}l@{}}0.021\\ (0.02)\end{tabular} & \textbf{\begin{tabular}[c]{@{}l@{}}0.007\\ (0.002)\end{tabular}} & \begin{tabular}[c]{@{}l@{}}0.021\\ (0.02)\end{tabular} & \begin{tabular}[c]{@{}l@{}}0.037\\ (0.01)\end{tabular} & \begin{tabular}[c]{@{}l@{}}0.01\\ (4e-3)\end{tabular}\\\bottomrule

\end{tabular}
\caption{Trading performance on S\&P 500.}
	\label{overall-result-us}

\end{table*}
\begin{table*}[htp]
\centering
\small
\begin{tabular}{@{}l||c|c|c|c|c||c|c@{}}
\toprule
Model &  & GGR & Cointegration & Correlation & Wang &  TRIALS & TRIALS wo TR \\ \midrule
\multirow{5}{*}{Set 1} & SR$\Uparrow$ & 0.22 & -2.19 & -0.20 & 0.19 & \textbf{0.75} & 0.73  \\\cmidrule{2-8}
 & AR$\Uparrow$ & 0.05 & -0.31 & -7e-3 & 0.05 & 0.13 & \textbf{0.14}  \\\cmidrule{2-8}
 & MDD$\Uparrow$ & -0.27 & -0.30 & -0.11 & -0.22 & \textbf{-0.08} & -0.16  \\\cmidrule{2-8}
 & AV$\Downarrow$ & 0.20 & 0.17 & \textbf{0.11} & 0.16 & 0.15 & 0.17 \\\cmidrule{2-8}
 & ED$\Downarrow$ & 0.007 & 0.008* & 0.011 & 0.008* & \textbf{0.006} & 0.015 \\ \midrule \midrule
\multirow{5}{*}{Set 2} & SR$\Uparrow$ & -1.18 & -1.55 & -1.27 & 0.17 & \textbf{2.41} & 1.11  \\\cmidrule{2-8}
 & AR$\Uparrow$ & -0.18 & -0.21 & -0.16 & 0.04 & \textbf{0.31} & 0.11 \\\cmidrule{2-8}
 & MDD$\Uparrow$ & -0.31 & -0.27 & -0.21 & -0.18 & -0.14 & -0.10 \\\cmidrule{2-8}
 & AV$\Downarrow$ & 0.17 & 0.16 & 0.14 & 0.12 & \textbf{0.10} & 0.19  \\\cmidrule{2-8}
 & ED$\Downarrow$ & 0.016 & \textbf{0.007} & 0.031 & 0.007 & 0.023 & 0.008  \\\midrule
\multirow{5}{*}{Set 3} & SR$\Uparrow$ & -1.54 & -0.60 & -0.62 & \textbf{1.61} & 0.31 & 0.56 \\\cmidrule{2-8}
 & AR$\Uparrow$ & -0.15 & -0.11 & -0.08 & \textbf{0.24} & 0.06 & 0.08 \\\cmidrule{2-8}
 & MDD$\Uparrow$ & -0.15 & -0.18 & -0.18 & -0.29 & \textbf{-0.12} & -0.12 \\\cmidrule{2-8}
 & AV$\Downarrow$ & 0.12 & 0.20 & 0.16 & \textbf{0.10} & 0.17 & 0.12 \\\cmidrule{2-8}
 & ED$\Downarrow$ & \textbf{0.007*} & 0.018 & 0.009 & 0.018 & 0.011 & 0.007 \\\midrule
\multirow{5}{*}{Set 4} & SR$\Uparrow$ & -1.90 & -2.68 & -1.06 & 0.79 & \textbf{2.57} & 2.36\\\cmidrule{2-8}
 & AR$\Uparrow$ & -0.29 & -0.47 & -0.22 & 0.29 & \textbf{0.53} & 0.45 \\\cmidrule{2-8}
 & MDD$\Uparrow$ & -0.08 & -0.10 & -0.07 & -0.70 & \textbf{0} & 0  \\\cmidrule{2-8}
 & AV$\Downarrow$ & 0.18 & 0.23 & 0.23 & 0.24 & 0.16 & \textbf{0.15} \\\cmidrule{2-8}
 & ED$\Downarrow$ & \textbf{0.013} & 0.028* & 0.014 & 0.028* & 0.056 & 0.018 \\\midrule
\multirow{5}{*}{Set 5} & SR$\Uparrow$ & -0.68 & -0.03 & -1.76 & \textbf{1.91} & 1.03 & 1.03 \\\cmidrule{2-8}
 & AR$\Uparrow$ & -0.14 & -6e-4 & -0.24 & \textbf{0.69} & 0.17 & 0.17 \\\cmidrule{2-8}
 & MDD$\Uparrow$ & -0.32 & -0.14 & -0.23 & -0.79 & -0.11 & -0.11 \\\cmidrule{2-8}
 & AV$\Downarrow$ & 0.22 & 0.18 & 0.16 & 0.20 & 0.15 & 0.15 \\\cmidrule{2-8}
 & ED$\Downarrow$ & 0.024 & 0.024 & \textbf{0.011} & 0.024 & 0.024 & 0.014 \\\midrule
\multirow{5}{*}{\begin{tabular}[c]{@{}l@{}}Mean\\ (Std)\end{tabular}} & SR$\Uparrow$ & \begin{tabular}[c]{@{}l@{}}-1.19\\ (0.74)\end{tabular} & \begin{tabular}[c]{@{}l@{}}-1.50\\ (0.97)\end{tabular} & \begin{tabular}[c]{@{}l@{}}-1.37\\ (0.25)\end{tabular} & \begin{tabular}[c]{@{}l@{}}0.75\\ (0.68)\end{tabular} & \textbf{\begin{tabular}[c]{@{}l@{}}1.91\\ (0.88)\end{tabular}} & \begin{tabular}[c]{@{}l@{}}0.95\\ (0.88)\end{tabular} \\\cmidrule{2-8}
 & AR$\Uparrow$ & \begin{tabular}[c]{@{}l@{}}-0.17\\ (0.11)\end{tabular} & \begin{tabular}[c]{@{}l@{}}-0.25\\ (0.17)\end{tabular} & \begin{tabular}[c]{@{}l@{}}-0.21\\ (0.07)\end{tabular} & \begin{tabular}[c]{@{}l@{}}0.24\\ (0.23)\end{tabular} & \textbf{\begin{tabular}[c]{@{}l@{}}0.68\\ (0.51)\end{tabular}} & \begin{tabular}[c]{@{}l@{}}0.13\\ (0.12)\end{tabular}\\\cmidrule{2-8}
 & MDD$\Uparrow$ & \begin{tabular}[c]{@{}l@{}}-0.29\\ (0.06)\end{tabular} & \begin{tabular}[c]{@{}l@{}}-0.29\\ (0.13)\end{tabular} & \begin{tabular}[c]{@{}l@{}}-0.25\\ (0.06)\end{tabular} & \begin{tabular}[c]{@{}l@{}}-0.18\\ (0.09)\end{tabular} & \begin{tabular}[c]{@{}l@{}}-0.14\\ (0.07)\end{tabular} & \textbf{\begin{tabular}[c]{@{}l@{}}-0.12\\ (0.09)\end{tabular}}\\\cmidrule{2-8}
 & AV$\Downarrow$ & \begin{tabular}[c]{@{}l@{}}0.18\\ (0.03)\end{tabular} & \begin{tabular}[c]{@{}l@{}}0.19\\ (0.03)\end{tabular} & \textbf{\begin{tabular}[c]{@{}l@{}}0.17\\ (0.05)\end{tabular}} & \begin{tabular}[c]{@{}l@{}}0.25\\ (0.07)\end{tabular} & \begin{tabular}[c]{@{}l@{}}0.26\\ (0.09)\end{tabular} & \begin{tabular}[c]{@{}l@{}}0.17\\ (0.07)\end{tabular} \\\cmidrule{2-8}
 & ED$\Downarrow$ & \textbf{\begin{tabular}[c]{@{}l@{}}0.013\\ (6e-3)\end{tabular}} & \begin{tabular}[c]{@{}l@{}}0.017\\ (8e-3)\end{tabular} & \begin{tabular}[c]{@{}l@{}}0.015\\ (8e-3)\end{tabular} & \begin{tabular}[c]{@{}l@{}}0.017\\ (8e-3)\end{tabular} & \begin{tabular}[c]{@{}l@{}}0.046\\ (0.02)\end{tabular} & \begin{tabular}[c]{@{}l@{}}0.02\\ (8e-3)\end{tabular} \\\bottomrule
\end{tabular}
\caption{Trading performance on CSI 300.}
	\label{overall-result-cn}
\end{table*}

\end{document}